\documentclass[12pt twoside]{article}
\usepackage{pifont}
\usepackage{verbatim}
\pdfoutput=1
\usepackage{tikz}
\usetikzlibrary{spy}
\usepackage{extarrows}
\usetikzlibrary{arrows,shapes}
\usepackage{cite}
\usepackage{amsmath,amssymb,latexsym,bm}
\usepackage{paralist}
\usepackage{graphicx}
\usepackage{float}
\usepackage{algorithmic}
\usepackage{algorithm}
\usepackage{cases}
\usepackage{subfigure}
\usepackage{caption2}
\usepackage{epstopdf}
\usepackage{makecell,multirow,diagbox}

\usepackage{xcolor}
\usepackage{tikz}
\usetikzlibrary{spy}
\usepackage{wrapfig}
\usetikzlibrary{arrows}
\tikzstyle{block}=[draw opacity=0.7,line width=1.4cm]
\tikzstyle{process} = [rectangle, minimum width=3cm, minimum height=1cm, text centered, draw=black, fill = yellow!50]

\usepackage[colorlinks=true]{hyperref}
\hypersetup{urlcolor=blue, citecolor=red,linkcolor=blue,}
\newcommand*{\affaddr}[1]{#1}
\newcommand*{\affmark}[1][*]{\textsuperscript{#1}}
\newcommand*{\email}[1]{\texttt{#1}}
\begin{document}
	\title{A Dataset-free Deep Learning Method for Low-Dose CT Image Reconstruction}
	\author{Qiaoqiao Ding\affmark[1],  Hui Ji\affmark[2], Yuhui Quan\affmark[3], and Xiaoqun Zhang\affmark[1,]\affmark[4]\\
		\affaddr{\affmark[1] Institute of Natural Sciences, Shanghai Jiao Tong University,\\ Shanghai 200240, China}\\
		\affaddr{\affmark[2] Department  of Mathematics, National University of Singapore,\\
			119076, Singapore}\\
		\affaddr{\affmark[3]  School of Computer Science and Engineering, \\South China University of Technology, Guangzhou 510006, China}\\
		\affaddr{\affmark[4]  School of Mathematical Sciences and MOE-LSC,\\ Shanghai Jiao Tong University, Shanghai 200240, China}\\	
		\email{\{dingqiaoqiao,xqzhang\}@sjtu.edu.cn}\\
		\email{matjh@nus.edu.sg}\\
		\email{csyhquan@scut.edu.cn}		
	}
	\maketitle
	\begin{abstract}
		Low-dose CT (LDCT) imaging  attracted a considerable interest  for the reduction of the object's exposure to X-ray radiation.
		In recent years, supervised deep learning (DL) has been extensively studied for LDCT image reconstruction, which  trains a  network over a dataset containing many pairs of normal-dose and low-dose images. However, 
		the challenge on collecting many such pairs in the clinical setup limits the application of such supervised-learning-based methods for LDCT image reconstruction in practice. Aiming at addressing the challenges raised by the collection of training dataset, this paper proposed  a unsupervised deep learning method for LDCT image reconstruction, which does not require any external training data. The proposed method is built on  a re-parametrization technique for  Bayesian inference via deep network with random weights, combined with additional total variational~(TV) regularization.
		The experiments show that the proposed method noticeably outperforms  existing dataset-free image reconstruction methods on the test data.
	\end{abstract}
	~~~~~~~	keywords: X-ray CT, Low Dose CT, Deep Neural Networks
	
	\section{Introduction}
	{X}{-ray} Computed Tomography (CT) has been widely applied in clinical imaging, for its ability of  providing high resolution images of internal anatomical structures.
	High-quality CT images are very useful for  prevention, diagnosis and treatments  of human diseases.  	However,  many studies indicated that excessive exposure to radiation  from X-ray CT scanner may be responsible for  the increasing risk of getting cancer, thus there is  the demand for reducing radiation  dose when run  CT  scanning.  There  are  two  main  techniques  for radiation  dose  reduction:
	decreasing the radiation exposure time, \emph{i.e.} decreasing the number of projection views~\cite{sidky2006accurate},  and lowering the X-ray tube current~\cite{whiting06pop}, \emph{i.e.}  LDCT. In comparison to  normal dose CT (NDCT), the signal-to-noise ratio (SNR) of the measurement in LDCT is much lower. As a result, the quality of the images reconstructed using the conventional methods is not satisfactory for LDCT, \emph{i.e.}, there are often noticeable streaky artifacts and random patterns appearing in the reconstruction image.
	
	The image reconstruction problem for LDCT  can be formulated as  solving a linear inverse problem 
	\begin{equation}\label{eqn:problem}
		\bm{y} = \bm{Ax} + \bm{n},
	\end{equation}
	where $\bm{A}$ denotes the projection matrix of CT imaging,  $\bm{y}$ denotes the available measurement, $\bm{x}$ denotes the image to be reconstructed, and 
	$\bm{n}$ denotes the measurement noise which is often modeled by i.i.d. random  variables. The inverse problem~\eqref{eqn:problem} from LDCT imaging is ill-posed. Certain regularization on $\bm{x}$ needs to be introduced to address solution ambiguity and to suppress noise magnification  when solving~\eqref{eqn:problem}.
	
	In recent years, deep learning (DL) has been a very promising tool for developing effective image reconstruction methods for CT, including LDCT. The majority of these existing deep learning solutions are based on supervised learning; see e.g. ~\cite{chen2017lowS,chen2017low,li2017low,jin2017deep,gupta2018cnn,adler2018learned,ding2020low,ding2021deep,he2018optimizing,Ye2018deep,ding2021learnable}. Supervised learning requires a dataset with many training samples,  the pairs of  the low-dose image and  normal dose image (assumed to be truth image). The construction of such image pairs requires two times scan and registration for alignment  of every image/projection pair, which is costly and troublesome in practice. In addition, the amount of real-world images for LDCT is very limited too. As a result, there  is  an  increasing  interest on  the  development  of  powerful  unsupervised  deep  learning methods  for  LDCT  imaging,  which  works  well  in  a  data-limited environment.
	\subsection{Related works}
	In existing literature,  many methods,  e.g., analytical filtering method~\cite{kak2002principles,hsieh2003computed,feldkamp1984practical} have been proposed to improve image quality of LDCT imaging. Due to low SNR of measurement data, these methods, including the ones equipped with  adaptive filtering~\cite{balda2012ray} and bilateral filtering~\cite{manduca2009projection},  are not capable of 
	produce high-quality CT images.   Over the past decades, iterative reconstruction algorithm is a popular approach 
	adopted in LDCT imaging, which is derived by minimizing a  cost function. The cost function usually is composed of a fidelity term determined by  statistical characters of noise and a regularization term induced by some pre-defined prior on the image. In the past, many regularizations have been proposed for LDCT image reconstruction, \emph{e.g.}, total variation (TV)~\cite{zhang2005total,sidky2008image}, wavelet tight frame-based sparsity prior~\cite{jia2011gpu}, nonlocal total variation~\cite{jia20104d},
	and low-rank  based patch prior~\cite{cai2014cine}.

	In past few years,  deep learning (DL)  has emerged as a prominent  tool for developing powerful  image reconstruction methods for LDCT imaging. 	 Earlier work on DL-based LDCT image reconstruction used  DL  as  a  post-processing  tool~\cite{chen2017lowS,chen2017low,li2017low,jin2017deep}, which trains a deep neural network (DNN) to denoise the images reconstructed from some existing works. The denoising network is  trained by using many pairs of images reconstructed from LDCT and the corresponding NDCT. Different network 
	architectures have been exploited in these works, \emph{e.g.},  convolution neural network (CNN)~\cite{chen2017lowS},  encoder-decoder CNN~\cite{chen2017low}, residual network~\cite{li2017low} and  U-Net~\cite{jin2017deep}. As the artifacts in the reconstructed images often cannot be modeled as independent random noise, the performance gain brought  by such a post process is limited. A more effective approach is the so-called optimization unrolling scheme~\cite{gupta2018cnn,adler2018learned,ding2020low,ding2021deep} and plug-\&-play \cite{he2018optimizing,Ye2018deepR}.  Such a scheme follows some iterative image reconstruction scheme derived from some regularization methods, and replaces the related regularization step by a learnable/pre-trained denoising DNN. The main difference among these methods lies in which iterative scheme is used for unrolling and how to train the  denoising network embedded inside the iterations.
	
	Recently, the development of dataset-free DL methods has drawn a lot of attention for LDCT imaging.  By using  a generative adversarial network (GAN), the works~\cite{wolterink2017generative,yang2018low}  trained the network on a dataset contains both low dose and normal dose images, which are not paired.  Inspired by recent works on unsupervised learning for generic image denoising, there are two approaches to extend these unsupervised denoising network to solve inverse problems such as low-dose CT image reconstruction.
	
	One unsupervised approach is treating CT image reconstruction as a denoising process which post-processes the reconstructed images. In \cite{hasan2020hybrid,yuan2020half2half}, the Noise2Noise (N2N)~\cite{lehtinen2018noise2noise}, a denoising network trained over the image pairs with independent noise, is introduced to remove artifacts of the reconstructed LDCT images. In
	\cite{hendriksen2020noise2inverse}, the Noise2Self (N2S)~\cite{batson2019noise2self}, an unsupervised denoising network trained over noisy images, is extended to solve the tomography problem by modeling artifacts of image as independent random noise.  Such an approach suffers from the same performance issue as its supervised counterparts that rely on a denoising network for  removing artifacts of the reconstructed images. The reason is that the artifacts of a reconstructed image are indeed highly correlated to the entries of the image, it cannot not well-modeled by simple random variables, such as i.i.d. noise, assumed by the unsupervised denoising networks. As a result, the performance of these unsupervised methods, derived from the unsupervised denoising networks above, is not very competitive to the state-of-the-art supervised methods for LDCT imaging.
	
	Another unsupervised approach is built on the so-called deep image prior (DIP)~\cite{ulyanov2018deep}. DIP is originally proposed for image denoising. It is empirically observed that when training a CNN to fit a noisy image, the regular image structures appear before random patterns. Thus, one can train a denoising network on a noisy image by early stopping. In other words, early stopping can be an effective technique for regularizing a denoising network. The DIP  has been exploited in various medical  imaging tasks, \emph{e.g. } such as PET
	reconstruction~\cite{gong2018pet,yokota2019dynamic}, MRI~\cite{yoo2021time},  diffraction tomography~\cite{zhou2020diffraction} and  compressed sensing~\cite{van2018compressed}. 	In \cite{baguer2020computed}, DIP is combined with TV-based regularization for CT image reconstruction. While DIP is simple and effective for image denoising, there are issues regarding DIP-derived methods for CT imaging. The artifacts in the reconstructed images are not random noise. They are also regular patterns correlated with the image structures. As a result, the early stopping adopted in DIP cannot prevent the appearance of artifacts in the reconstructed image.

	\subsection{Our approach}
	In this paper, we present an unsupervised deep learning method for LDCT image reconstruction, without requiring any external training samples with truth images. Such an unsupervised method certainly can see its great value in practice. The proposed method is built on the Bayesian inference where the prior distribution of an image is re-parametrized by a deep network with random weights. 
	
	Recall that in Bayesian inference, we have two representative Bayesian estimators. One is the maximum a posterior (MAP) estimator:
	\begin{align}
		\bm{x}_{_{\text{MAP}}}=&\arg\max_{\bm{x}} p(\bm{x}|\bm{y}),
		\label{eq:MAP}
	\end{align}
	and the other is the minimum mean squared error estimator (MMSE) estimator or equivalently conditional mean estimator
	\begin{align}
		\bm{x}_{_{\text{CM}}}&=\mathbb{E}_{(\bm{x}|\bm{y})}(\bm{x}|\bm{y})
		=\int \bm{x} p(\bm{x}|\bm{y}) d\bm{x},
		\label{eq:CM}
	\end{align}
	where  $ p(\bm{x}|\bm{y})$ denote the posterior distribution of $\bm{x}$ given the measurement $\bm{y}$. The key to both estimators is about deriving the posterior distribution $ p(\bm{x}|\bm{y})$ which models the data well. A common practice in Bayesian inference is to re-express
	$ p(\bm{x}|\bm{y})$ by Bayesian rule:
	$$
	p(\bm{x}|\bm{y})=p(\bm{y}|\bm{x})p(\bm{x})/p(\bm{y}),
	$$
	where the likelihood term $p(\bm{y}|\bm{x})$ can be expressed as 
	$$
	p(\bm{y}|\bm{x})=\frac{1}{2\sigma^2}\|\bm{y}-\bm{A}\bm{x}\|_2^2
	$$
	in the presence of i.i.d Gaussian white noise $\bm{n}\sim \mathcal{N}(0,\sigma^2 \bm{I})$. Then, the study of the estimators turns to defining a prior distribution $p(\bm{x})$ that accurately models statistical characters of images for reconstruction.
	
	In traditional regularization methods, to be computationally tractable, the prior distribution $p(\bm{x})$  usually is modeled by mean-field approximation which assumes the independence of all image  pixels.  For example, the well-know TV-based regularization assumes 
	$$
	p(\bm{x})=\prod_i q((\nabla x)_i),
	$$
	where $q$ is the density function of a Laplacian distribution: $q(z)\sim e^{-\frac{|z|}{\lambda}}$.
	There are two concerns in the assumptions of the prior distribution used in TV-based regularization. One is the mean-field assumption and the other is the Laplacian assumption.
	
	The over-simplified mean-field model for the prior distribution $p(\bm{x})$ motivates us to study a different approach to model the prior distribution $p(\bm{x})$ or the posterior distribution $p(\bm{x}|\bm{y})$.
	Inspired by the advance of optimization technique for solving the optimization problems of network training, we proposed to adopt a   re-parametrization technique for Bayesian inference, which re-express the variable $\bm{x}$ by a DNN with random weights
	$$
	\bm{x}=f(\bm{x}_0;\bm{\theta}),
	$$
	where $\bm{x}_0$ is some initial seed and $\bm{\theta}$ are random variables.
	It can be seen that after re-parametrization, the prior distribution of $p(\bm{x})$ can be very complicated, even though the variable $\bm{\theta}$ is modeled by mean-field approximation.
	
	After re-parametrization, the variables for inference now  are  random network weights $\bm{\theta}$. Again, the key for a Bayesian inference now is to define an appropriate posterior distribution $p(\bm{\theta}|\bm{y})$ for $\bm{\theta}$.
	As in general, it is  not computational tractable in high dimension, we adopt the variational approximation method which approximate
	$p(\bm{\theta}|\bm{y})$ by a set of approximation distributions $q(\bm{\theta}| \bm{\mu})$ parametrized by $\bm{\mu}$. The optimal approximation with distribution parameters $\bm{\mu}^\ast$ is then estimated by minimizing the Kullback–Leibler (KL) divergence between two  distributions. Once the approximation to posterior distribution $
	p(\bm{\theta}|\bm{y})
	$ is obtained, we can utilize the Bayesian inference to estimate the image.

	This paper is organized as follows.  
	Section~\ref{Method}  describes the proposed method and algorithm.
	The NN architecture and implementation details of NN are given in  Section~\ref{Implementation}.
	Section~\ref{Experiments} is devoted to the experimental evaluation and  comparison to other methods.
	Section~\ref{conclusion} concludes the paper.
	
	\section{Method}
	\label{Method}
	In this section, we  give a detailed discussion on the proposed self-supervised method for LDCT reconstruction from noisy measurement, which is built on the DNN-based re-parametrization for Bayesian inference.
	Recall that CT reconstruction problems can be formulated as the following inverse problem: given an observed image $\bm{y}\in \mathbb{R}^m$ corrupted according to forward model and noise, $\bm{n}$, find the unknown image $\bm{x}\in \mathbb{R}^n$ which satisfies the observation
	\begin{eqnarray}
		\bm{y}=\bm{Ax}+\bm{n}. \label{eq:problem}
	\end{eqnarray}
	
	Considering a DNN with random weights for the re-parametrization: 
	$$
	\bm x=f(\bm{x}_0;\bm{\theta}).
	$$
	Then,  the inference of $\bm{x}$ from noisy measurement $\bm{y}$ is now about inferring the network weights $\bm{\theta}$ from $\bm{y}$. In order to perform Bayesian inference for $\bm{\theta}$, the key is to derive the posterior distribution $p(\bm{\theta}|\bm{y})
	$. As  $p(\bm{\theta}|\bm{y})
	$ is in general computationally intractable, we propose to approximate it by the following set of distributions $q(\bm{\theta}| \bm{\mu})$ defined by
	\begin{equation}
		\label{eq:theta}
		\bm{\theta}= \bm{\mu} \odot \bm{b}:\quad \theta_i= \mu_i\cdot b_i, 1\leq i\leq N,
	\end{equation}
	where  $\mu_i$ denotes the distribution parameter of $\theta_i$ and $b_i\sim \mathbf{B}(p_i)$ follows a Bernoulli distribution with probability $p_i$. In other words, 
	the probability density  function of $b_i$   is defined as
	\begin{align}
		p(b_i)=p_i^{b_i}(1-p_i)^{1-b_i} ~~~~~~~b_i=\{0,1\}.	
	\end{align}
	In other words, the DNN with random weights used in this  paper is the widely used the network with dropout.  It is noted that the idea of using the network with dropout also has been exploited in S2S~\cite{quan2020self2self,chen2022nonblind,li2022supervised} for  self-supervised image denoising and deconvolution.
	
	In the next, we give a detailed discussion on how to train the network by minimizing the KL divergence between $q(\bm{\theta}|\bm{\mu})$, and how to use the trained model for testing by using Monte-Carlo(MC) sampling.
	
	\subsection{Training}\label{sec:train}
	As we use $q(\bm{\theta}| \bm{\mu})$ to approximate
	$p(\bm{\theta}|\bm{y})$, the optimal approximation is estimated by 
	minimizing the KL-divergence between $q(\bm{\theta}| \bm{\mu})$  and $p(\bm{\theta}|\bm{y})$:
	\begin{align}
		&\min_{\bm{\mu}} {\rm{KL}}(q(\bm{\theta}| \bm{\mu})||p(\bm{\theta}|\bm{y}))\nonumber\\
		=&\min_{\bm{u}}  \mathbb{E}_{\bm{\theta}\sim q(\bm{\theta}| \bm{\mu})}  [\log q(\bm{\theta}| \bm{\mu})- \log p(\bm{\theta}|\bm{y})] \nonumber\\
		\propto&\min_{\bm{\mu}}  \mathbb{E}_{\bm{\theta}\sim q(\bm{\theta}| \bm{\mu})}  [\log q(\bm{\theta}| \bm{\mu})- \left(\log p(\bm{y}|\bm{\theta})+\log p(\bm{\theta})\right)] \nonumber\\
		=&\min_{\bm{\mu}}{\rm{KL}}(q(\bm{\theta}| \bm{\mu})||p(\bm{\theta}))-\mathbb{E}_{\bm{\theta}\sim q(\bm{\theta}| \bm{\mu})}\log p(\bm{y}|\bm{\theta}).
	\end{align}	
	For the first term, suppose that $p(\bm{\theta})$ is a uniform distribution in a sufficient larger region $\Omega$.
	Here, we abuse the notion $\frac{0}{0}=1$. We have	
	$q(\theta_i| \mu_i)=p_i^{\frac{\theta_i}{\mu_i}}(1-p_i)^{1-\frac{\theta_i}{\mu_i}}, ~~\theta_i=\{0,\mu_i\}$ and
	$p(\theta_i)=1/s_i$, where $s_i$ is the length of the domain of definition about $\theta_i$.
	Then,
	\begin{align}
		D_{KL}(q(\bm{\theta}| \bm{\mu})||p(\bm{\theta}))&=\sum_i D_{KL}(q(\theta_i| \mu_i )||p(\theta_i)),\nonumber\\
		&=\sum_i   q(\theta_i| \mu_i )\log{\frac{q(\theta_i| \mu_i )}{ p(\theta_i)}},\nonumber\\
		&=\sum_i (1-p_i)\log{(1-p_i) }+p_i\log{p_i }+\log s_i.\nonumber
	\end{align}
	
	Finally, we obtain
	\begin{align}
		D_{KL}(q(\bm{\theta}| \bm{\mu})||p(\bm{\theta}))=c_0,\quad \bm{\theta}\in\Omega,
	\end{align}
	where $c_0$ is a constant.
	
	In the second term, 	    suppose that the measurement noise $\bm{n}$ is Gaussian white noise
	such that $p(\bm{n})\propto\prod_i \exp(\frac{-n_i^2}{2\tilde{\sigma}^2})$, we have
	$$\log(p(\bm{y}|\bm{\theta}))\propto-\frac{1}{2\tilde{\sigma}^2}\| \bm{A}f(\bm{x}_0,{\bm{\theta}})-\bm{y}\|_2^2.$$
	Then, we have
	\begin{align}
		\min_{\bm{\mu}} {D_{KL}}(q(\bm{\theta}| \bm{\mu})||p(\bm{\theta}|\bm{y}))
		\propto\min_{\bm{\mu}} \mathbb{E}_{\bm{\theta}\sim q(\bm{\theta}| \bm{\mu})}\| \bm{A}f(\bm{x}_0,{\bm{\theta}})-\bm{y}\|_2^2.
		\label{eqn:KL}
	\end{align}

	It can be seen from \eqref{eqn:KL} that the KL divergence only constrains the estimation in the range space of the projection matrix $\bm{A}$. To avoid  possible overfitting, we introduce an additional regularization on the estimation, and we adopt the widely-used TV regularization to  the loss function. 
	Recall  \eqref{eq:theta} and consider the definition of $q(\bm{\theta}| \bm{\mu})$ and $\bm{B}(\bm{p})$,  we deduce that 
	\begin{align}
		\min_{\bm{\mu}}&\mathbb{E}_{\bm{\theta}\sim q(\bm{\theta}| \bm{\mu})}\| \bm{A}f(\bm{x}_0,\bm{\theta})-\bm{y}\|_2^2\nonumber\\
		&=\min_{\bm{\mu}}\int \|\bm{A}f(\bm{x}_0,\bm{\theta})-\bm{y}\|_2^2~q(\bm{\theta}| \bm{\mu}) d\bm{\theta}\nonumber\\
		&\xlongequal[]{d\bm{\theta}=\bm{\mu} \odot d\bm{b}}\min_{\bm{\mu}}\int \|\bm{A}f(\bm{x}_0,\bm{\mu}\odot\bm{b})-\bm{y}\|_2^2~ \bm{B}(\bm{p}) d\bm{b}\nonumber\\
		&=\min_{\bm{\mu}} \mathbb{E}_{\bm{b}\sim \bm{B}(\bm{p})}\| \bm{A}f (\bm{x}_0,\bm{\mu}\odot\bm{b})-\bm{y}\|_2^2\nonumber.
	\end{align}
	The final loss function for training the network now is 
	\begin{align}
		\min_{\bm{\mu}} \mathbb{E}_{\bm{b}\sim \bm{B}(\bm{p})}\| \bm{A}f(\bm{x}_0,\bm{\mu}\odot\bm{b})-\bm{y}\|_2^2+\alpha \|\nabla f(\bm{x}_0,\bm{\mu}\odot\bm{b})\|_1,
		\label{TrainModel}
	\end{align}  
	where $\alpha$ is a pre-defined hyper-parameter.
	When training the network, the loss function is minimized by using  MC dropout \cite{gal2016dropout}, \emph{i.e.}, randomly  dropping out nodes during the training with dropout rate $\bm{p}$.

	\subsection{Testing}\label{sec:test}
	Once the NN is trained via minimizing the loss function given in 
	\eqref{TrainModel}, we have an approximation to the posterior distribution $p(\bm{\theta}|\bm{y})$, denoted by $q(\bm{\theta}|\bm{\mu^\ast})$. In our approach, we estimate the image $\bm{x}$ using conditional mean estimator. Recall that given the measurement $\bm{y}$, its conditional mean estimator for $\bm{x}$ reads 
	$$\bm{x}_{_\text{CM}}=\int \bm {x}p (\bm{x}|\bm{y})d\bm{x}.$$
	By re-parametrization: $\bm x=f(\bm{x}_0;\bm{\theta})$,  we have
	$$\bm{x}_{_\text{CM}}=\int \bm {x}p (\bm{x}|\bm{y})d\bm{x}=\int f(\bm{x}_0;\bm{\theta})p(\bm{\theta}|\bm{y})d\bm{\theta}.$$
	By approximating $p(\bm{y}|\bm{\theta})$ using $q(\bm{\theta}|\bm{\mu^\ast})$, we have an approximate conditional mean estimator of $\bm x$ given by
	\begin{align}
		\bm{x}^\ast_{_\text{CM}}=
		\int f(\bm{x}_0;\bm{\theta}|\bm{\mu}^\ast)q(\bm{\theta}|\bm{\mu}^\ast) d\bm{\theta}.
	\end{align}
	The integration above is calculated by using  MC integration in practice. That is, after the network is trained, we take 
	${K}$ random samples of the networks with dropout:
	$$
	f(\bm{x}_0;\bm{\theta}_k)=f(\bm{x}_0;\bm{\mu}^\ast\odot{\bm b_k}),\quad
	\bm{b}_k\sim \bm {B}(\bm{p}).
	$$
	Then, the estimate is defined by taking the the average of these $K$ samples:
	\begin{align}
		\bm{x}^\ast=\frac{1}{{K}}\sum_{k=1}^{{K}}f(\bm{x}_0;\bm{\theta}_k)=\frac{1}{{K}}\sum_{k=1}^{{K}}f(\bm{x}_0;\bm{\mu}^\ast\odot{\bm b_k}).
		\label{MCInference}
	\end{align}
	
	\subsection{Discussion}
	In sections~\ref{sec:train} and~\ref{sec:test}, we present a DNN-based re-parametrization $\bm{x}=f(\bm{x}_0;\bm{\theta})$ for facilitating the Bayesian inference of LDCT image reconstruction. In the proposed approach, the corresponding posterior distribution $p(\bm{\theta}|\bm{y})$ is approximated by a network with dropout $q(\bm{\theta}|\bm{\mu})$ via minimizing their KL divergence. After the network is trained with dropout. The network is sampled with drop-out to have a MC-based approximation to the conditional mean estimator of $\bm{x}$. 
	
	In addition, as only the noisy measurement $\bm{y}$ is available which only measures the image $\bm{x}$ in the range space of $\bm{A}$, a TV-regularization is introduced in the loss function for regularizing the network to avoid possible overfitting.  As a result, the loss function \eqref{TrainModel} is closely connected to the non-learning TV regularization method for solving inverse problems. Indeed, based on the loss function \eqref{TrainModel}, the proposed method can be viewed as learning multiple solvers to the TV-regularization model, and each solver differentiate itself from others by using different network architectures (by random dropout). From the perspective of ensemble learning, the proposed method can be also interpreted as an ensemble learning method that is built on TV-related regularization. It is likely that the artifacts from each instance of the solver to TV regularization have certain degree of independence. Then, the average of the results from these solvers will benefit such artifact independence to have an estimate with less artifacts. In short, the proposed method provides an efficient ensemble learning method for LDCT image via dropout.
	\subsection{Implementation details}
	\label{Implementation}
	
	\subsubsection{NN architecture}
	To evaluate the effectiveness of the proposed method, we test it using an encoder-decoder with skip-connection as the backbone network, whose diagram is illustrated in
	Fig.~\ref{fig:1}~(a). In the diagram of the network, the notation $D_i$, $U_i$ and $S_i$ represent the downsampling, upsampling and skip-connection blocks in the NN respectively.
	In the decoder-encoder architecture, $c_u[i]$, $c_d[i]$, $c_s[i]$ correspond to the number of filters at depth $i$ for
	the upsampling, downsampling, skip-connections respectively. The values $k_u[i]$, $k_d[i]$, $k_s[i]$ correspond to the respective
	kernel sizes. The values  $p_u[i]$, $p_d[i]$, $p_s[i]$ are the drop probability for
	the upsampling, downsampling, skip-connections respectively. Note that there is no upsampling layer in $U_1$ and the NN structure is similar to a U-Net.

	\tikzstyle{Block} = [rectangle,draw=cyan, fill = cyan!50, rotate=-90]
	\tikzstyle{arrow} = [single arrow, align=center,fill=lime!20, draw=teal, minimum height=3cm, minimum width=0.5cm]
	\tikzstyle{circ} = [circle,scale=0.5, draw=black, fill = black]
	\tikzstyle{arrowSkip} = [single arrow, fill=orange!20, draw=orange, align=center, rotate=-90,  minimum height=1.5cm, minimum width=0.5cm]
	\tikzstyle{arrowup} = [single arrow, fill=red!20, draw=red, align=center, rotate=180,  minimum height=3cm, minimum width=0.5cm]
	\tikzstyle{Format} = [draw, thin, draw =white,fill=white]
	
	\tikzstyle{BlockConv} = [rectangle, minimum width=0.1cm, minimum height=4cm, text centered, draw=teal, fill = teal!60]
	\tikzstyle{BlockDown} = [rectangle, minimum width=0.1cm, minimum height=3cm, text centered, draw=gray, fill =gray!70]
	\tikzstyle{BlockBatch} = [rectangle, minimum width=0.1cm, minimum height=3cm, text centered, draw=orange, fill = orange!50]
	\tikzstyle{BlockRelu} = [rectangle, minimum width=0.1cm, minimum height=3cm, text centered, draw=red, fill = red!50]
	\tikzstyle{BlockUp} = [rectangle, minimum width=0.1cm, minimum height=3cm, text centered, draw=purple, fill = purple!50]
	
	\tikzstyle{BlockDrop} = [rectangle, minimum width=0.1cm, minimum height=3cm, text centered, draw=brown, fill = brown!50]
	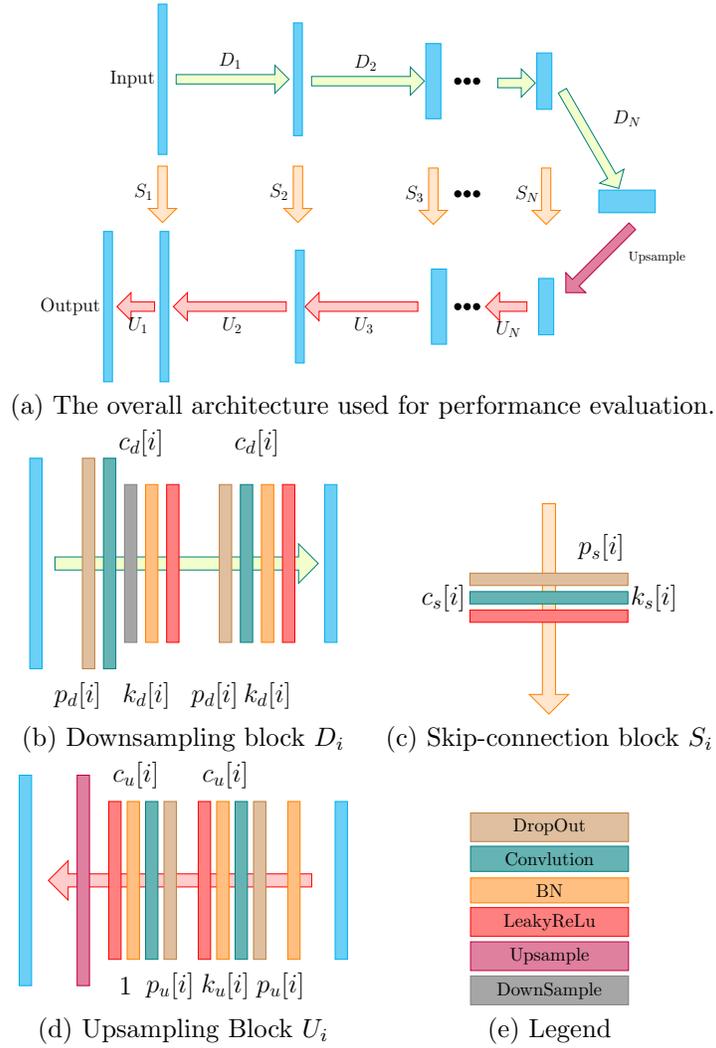
\begin{figure}[!htp]
		\centering
		\scalebox{0.5}{
			\begin{tikzpicture}[node distance=0.5cm]
				\path[->] node[Block, minimum width=4.cm, minimum height=.1cm] (Input)at (0,0) {};
				\path[->] node [left of=Input,black,xshift=-0.3cm,yshift=0.0cm] {\Large Input};
				\path[->] node[arrow, right of=Input,  xshift=1.3cm, yshift=-0.0cm](d1){ };
				\path[->] node[above of=d1]{\Large$D_1$ };
				\path[->] node[Block,right of=d1, xshift=-0.5cm,yshift=1.8cm, minimum width=3.cm, minimum height=.2cm] (I1) {};
				\path[->] node[arrow,right of=I1, xshift=1.3cm, yshift=-0.05cm](d2){ };
				\path[->] node[above of=d2]{\Large$D_2$ };
				\path[->] node[Block,right of=d2, xshift=-0.5cm,yshift=1.8cm ,  minimum width=2.cm, minimum height=.4cm] (I2) {};
				\path[->] node[circ,right of=I2,  xshift=0.8cm] (dot1){};
				\path[->] node[circ,right of=dot1,xshift=-0.0cm] (dot2){};
				\path[->] node[circ,right of=dot2,xshift=-0.0cm] (dot3){};
				\path[->] node[arrow, right of=dot3, xshift=0.5cm, yshift=-0.05cm, minimum height=1.0cm](d3){ };
				\path[->] node[arrow, right of=dot3, xshift=2.5cm, yshift=-1.5cm, rotate=-60](d4){ };
				\path[->] node[above of=d4,xshift=1.0cm]{\Large$D_N$ };
				\path[->] node[Block,right of=dot3, xshift=-0.5cm,yshift=1.8cm ,  minimum width=1.5cm, minimum height=.4cm] (I3) {};
				\path[->] node[Block,right of=d4, xshift=1.2cm,yshift=1.0cm, minimum width=.6cm, minimum height=1.5cm] (I4) {};
				\path[->] node[arrowSkip, below of=Input,xshift=3.0cm, yshift=0.5cm](s0){ };
				\path[->] node[left of=s0]{\Large$S_1$ };
				\path[->] node[arrowSkip, below of=I1,xshift=3.0cm, yshift=0.5cm](s1){ };
				\path[->] node[left of=s1]{\Large$S_2$ };
				\path[->] node[arrowSkip, below of=I2,xshift=3.0cm, yshift=0.5cm](s2){ };
				\path[->] node[left of=s2]{\Large$S_3$ };
				\path[->] node[arrowSkip, below of=I2,xshift=3.0cm, yshift=3.5cm](sN){ };
				\path[->] node[left of=sN]{\Large$S_N$ };
				\path[->] node[arrowSkip, below of=I4,xshift=1.5cm, yshift=-0.2cm,minimum height=2.5cm,rotate=-45,draw=purple, fill = purple!50](up){ };
				\path[->] node[right of=up,xshift=1.0cm]{Upsample };
				\path[->] node[circ,below of=dot1,yshift=-5.5cm,xshift=0.0cm] (dots1){};
				\path[->] node[circ,below of=dot2,yshift=-5.5cm,xshift=-0.0cm] (dots2){};
				\path[->] node[circ,below of=dot3,yshift=-5.5cm,xshift=-0.0cm] (dots3){};
				\path[->] node[Block, xshift=3cm,yshift=0.5cm,below of=sN, minimum width=1.5cm, minimum height=.4cm] (o4) {};
				\path[->] node[arrowup, left of=o4,  xshift=1.5cm, yshift=-0.0cm, minimum height=1.1cm](u3){ };
				\path[->] node[below of=u3,yshift=-0.1cm]{\Large$U_N$ };
				\path[->] node[circ,below of=dots1,yshift=-5.5cm,xshift=0.0cm] (dotss1){};
				\path[->] node[circ,below of=dots2,yshift=-5.5cm,xshift=-0.0cm] (dotss2){};
				\path[->] node[circ,below of=dots3,yshift=-5.5cm,xshift=-0.0cm] (dotss3){};
				\path[->] node[Block, xshift=0.5cm,yshift=-1.0cm,left of=dotss3, minimum width=2cm, minimum height=.4cm] (o3) {};
				\path[->] node[arrowup, left of=o3,  xshift=2.5cm, yshift=-0.0cm](u2){ };
				\path[->] node[below of=u2]{\Large$U_3$ };
				\path[->] node[Block, xshift=0.5cm,yshift=-1.7cm,left of=u2, minimum width=3cm, minimum height=.2cm] (o2) {};
				\path[->] node[arrowup, left of=o2, xshift=2.3cm, yshift=-0.0cm](u1){ };
				\path[->] node[below of=u1]{\Large$U_2$ };
				\path[->] node[Block,  left of=u1,xshift=0.5cm,yshift=-1.8cm, minimum width=4cm, minimum height=0.1cm] (o1) {};
				\path[->] node[arrowup, left of=o1,  xshift=1.2cm, yshift=-0.0cm,minimum height=1cm, minimum width=0.5cm](u0){ };
				\path[->] node[below of=u0]{\Large$U_1$ };
				\path[->] node[Block, left of=u0,xshift=0.5cm,yshift=-0.8cm, minimum width=4cm, minimum height=0.1cm] (oo) {};
				\path[->] node[right of=oo,xshift=-1.5cm]{\Large Output };
			\end{tikzpicture}
		}\\
		(a)  The overall architecture used for performance evaluation.\\
		\begin{tabular}{cc}
			\scalebox{0.7}{\begin{tikzpicture}
					[node distance=0.5cm]
					\path[->] node[Block, minimum width=4.cm, minimum height=.1cm] (Input)at (0,0) {};
					\path[->] node[arrow, minimum height=5cm,right of=Input,  xshift=2.3cm, yshift=-0.0cm](down){ };
					\path[->] node[BlockDrop,right of=Input,minimum height=4cm,xshift=0.5cm ] (Drop1){};
					\path[->] node[BlockConv,right of=Drop1,xshift=-0.1cm] (Conv1){};
					\path[->] node[BlockDown,right of=Conv1,xshift=-0.1cm] (Down1){};
					\path[->] node[BlockBatch,right of=Down1,xshift=-0.1cm] (Batch1){};
					\path[->] node[BlockRelu,right of=Batch1,xshift=-0.1cm] (Relu1){};
					\path[->] node[above of=Down1,yshift=1.8cm,xshift=0.2cm]{\Large$c_d[i]$ };
					\path[->] node[below of=Drop1,yshift=-2cm,xshift=-0.2cm]{\Large$p_d[i]$ };
					\path[->] node[below of=Down1,yshift=-2cm,xshift=0.3cm]{\Large$k_d[i]$ };
					\path[->] node[BlockDrop,right of=Down1,minimum height=3cm,xshift=1.3cm ] (Drop2){};
					\path[->] node[BlockConv,right of=Drop2,xshift=-0.1cm, minimum height=3cm] (Conv2){};
					\path[->] node[BlockBatch,right of=Conv2,xshift=-0.1cm] (Batch2){};
					\path[->] node[BlockRelu,right of=Batch2,xshift=-0.1cm] (Relu2){};
					\path[->] node[above of=Conv2,yshift=1.8cm,xshift=0.2cm]{\Large$c_d[i]$ };
					\path[->] node[below of=Conv2,yshift=-2cm,xshift=-0.6cm]{\Large$p_d[i]$ };
					\path[->] node[below of=Conv2,yshift=-2cm,xshift=0.4cm]{\Large$k_d[i]$ };
					\path[->] node[Block,right of=down, xshift=-0.5cm,yshift=2.8cm, minimum width=3.cm, minimum height=.2cm] {};
			\end{tikzpicture}} &
			\scalebox{0.7}{
				\begin{tikzpicture}
					\path[->] node[arrowSkip,minimum height=4.0cm](skip)at (0,0){ };
					\path[->] node[BlockDrop,below of=skip,xshift=0.0cm,yshift=1.5cm,rotate=-90 ] (Drop1){};
					\path[->] node[BlockConv,below  of=Drop1,xshift=0.0cm,yshift=0.65cm, minimum height=3cm,rotate=-90] (Conv1){};
					\path[->] node[BlockRelu,below  of=Conv1,xshift=0.0cm,yshift=0.65cm,rotate=-90] (Relu1){};
					\path[->] node[left of=Conv1,yshift=0.0cm,xshift=-1cm]{\Large$c_s[i]$ };
					\path[->] node[right of=Conv1,yshift=-0.0cm,xshift=1cm]{\Large $k_s[i]$ };
					\path[->] node[above of=Drop1,yshift=-0.4cm,xshift=1cm]{\Large $p_s[i]$ };
				\end{tikzpicture}
			}\\
			(b) Downsampling block $D_i$&(c) Skip-connection block $S_i$\\
			\scalebox{0.7}{
				\begin{tikzpicture}
					\path[->] node[Block, minimum width=3.cm, minimum height=.1cm] (Input)at (0,0) {};
					\path[->] node[arrowup, left of=Input, minimum height=5cm,  xshift=4.0cm, yshift=-0.0cm](up){ };
					\path[->] node[BlockBatch, left of=Input,  xshift=0.1cm, yshift=-0.0cm](BN1){ };
					
					\path[->] node[BlockDrop,left of=BN1,minimum height=3cm,xshift=0.35cm ] (Drop1){};
					\path[->] node[BlockConv,left  of=Drop1,xshift=0.65cm, minimum height=3cm] (Conv1){};

					\path[->] node[BlockBatch,left  of=Conv1,xshift=0.65cm] (Batch1){};
					\path[->] node[BlockRelu,left  of=Batch1,xshift=0.65cm] (Relu1){};
					\path[->] node[above of=Relu1,yshift=1.0cm,xshift=0.4cm]{\Large$c_u[i]$ };
					\path[->] node[below of=Relu1,yshift=-1.0cm,xshift=0.4cm]{\Large$k_u[i]$ };
					\path[->] node[below of=Drop1,yshift=-1.0cm,xshift=0.4cm]{\Large$p_u[i]$ };
					\path[->] node[BlockDrop,left of=Relu1,minimum height=3cm,xshift=0.35cm ] (Drop2){};
					\path[->] node[BlockConv,left  of=Drop2,xshift=0.65cm, minimum height=3cm] (Conv2){};
					\path[->] node[BlockBatch,left  of=Conv2,xshift=0.65cm] (Batch2){};
					\path[->] node[BlockRelu,left  of=Batch2,xshift=0.65cm] (Relu2){};
					\path[->] node[above of=Relu2,yshift=1.0cm,xshift=0.4cm]{\Large$c_u[i]$ };
					\path[->] node[below of=Relu2,yshift=-1.0cm,xshift=0.2cm]{\Large 1 };
					\path[->] node[BlockUp,left  of=Relu2,xshift=0.4cm, minimum height=4cm] (Up1){};
					\path[->] node[below of=Drop2,yshift=-1.0cm,xshift=0.0cm]{\Large$p_u[i]$ };
					\path[->] node[Block, minimum width=4.cm,left of=up,yshift=-3.0cm,xshift=1.0cm, minimum height=.1cm]  {};
				\end{tikzpicture}
			}
			&
			\scalebox{0.7}{
				\begin{tikzpicture}
					\path[->] node[BlockDrop, minimum width=3.cm, minimum height=.1cm] (Drop1){DropOut};
					\path[->] node[BlockConv,below of=Drop1,xshift=0.0cm,yshift=0.4cm, minimum width=3.cm, minimum height=.1cm] (Conv1){Convlution};
					\path[->] node[BlockBatch,below of=Conv1,xshift=0.0cm,yshift=0.4cm, minimum width=3.cm, minimum height=.1cm] (Batch1){BN};
					\path[->] node[BlockRelu,below of=Batch1,xshift=0.0cm,yshift=0.4cm, minimum width=3.cm, minimum height=.1cm] (Relu1){LeakyReLu};
					\path[->] node[BlockUp,below of=Relu1,xshift=0.0cm,yshift=0.35cm, minimum width=3.cm, minimum height=.1cm] (Up1){Upsample};
					\path[->] node[BlockDown,below of=Up1,xshift=0.0cm,yshift=0.35cm, minimum width=3.cm, minimum height=.1cm] (Down1){DownSample};
				\end{tikzpicture}
			} \\
			(d) Upsampling Block $U_i$&(e) Legend
		\end{tabular}
		\caption{
			Diagram of the network used for evaluating the proposed method.
		}
		\label{fig:1}
	\end{figure}
	\subsubsection{Implementation}
	For the implementation of the network, 
	the number of layer $N$ is set to $5$. For the layers from $i=1,\cdots, N$, the filter numbers are set as $c_d[i]=c_u[i]=128$  and $c_s[i]=4$.  Conv layers are with kernel size of $k_d[i]=k_u[i]=3$ and $k_s[i]=1$, strides of 1, and reflection padding of length 2 with $i=1,\cdots, N$.  LeakyReLU \cite{he2015delving} is used as  the non-linear activation unit where the slop  is set to 0.1. 
	Max pooling is used for downsampling, and the bi-linear interpolation is used for upsampling.  There is  no dropout in downsampling and upsampling block, \emph{i.e.}  the dropout probability of $D_i$ and $U_i$ are set to 0.
	For the other blocks, the dropout is conducted element-wisely with dropout probability  set to $p_s[i]=0.3$.
	
	For the initial value $\bm{x}_0$, we adopt the $\mathcal{J}$-invariant transform of the FBP reconstructed image $\bm{x}_{_{\bf{FBP}}}$ as  \cite{batson2019noise2self},
	$$	\bm{x}_0=\bm{b}\odot\bm{x}_{_{\bf{FBP}}}+(1-\bm{b})\odot \bm{s}(\bm{x}_{_{\bf{FBP}}}),$$
	where  $\odot$ denotes the element-wise multiplication,  
	$\bm{b}$ denotes a binary Bernoulli vector whose entries are independently sampled from a Bernoulli distribution with probability $p=0.3$,
	and the function $\bm{s}(\cdot)$ is set to the convolution with kernel  $\frac{1}{6}[\frac{1}{2}, 1, \frac{1}{2}; 1, 0, 1;\frac{1}{2}, 1, \frac{1}{2}]$. Note that such an initialization can be implemented by adding a Conv Layer with enabled dropout and a pre-defined low-pass filter.
	For the MC sampling in testing, we set $K=50$.
	
	\section{Experiments}
	\label{Experiments}
	The proposed method is implemented in PyTorch \cite{paszke2017automatic} interface on a NVIDIA Titan GPU. Adam optimizer is used with the momentum parameter $\beta=0.9$, and the learning rate is set as $10^{-5}$.
	\subsection{Methods in comparison}
	To evaluate its performance on LDCT image reconstruction, the proposed method is compared to  several  representative non-learning  methods, including  TV-based penalized weighed least squares method (PWLS-TV), KSVD \cite{zhang2010discriminative} and BM3D \cite{dabov2007image}. 		
	The PWLS-TV method uses the following regularization model for reconstructing the image from the measurement: 
	\begin{align}
		\arg\min_{\bm{x}} \|\bm{Ax}-\bm{y}\|_2^2+\alpha\|\nabla\bm{x}\|_1, \label{PWLSTV}
	\end{align}
	where $\alpha$ is regularization parameter. It is solved by the ADMM method in our experiments.

	Also, the proposed method is  compared to two recent unsupervised learning methods and a supervised learning method. Two unsupervised learning methods include 
	S2S for denoising-based post-process and DIP+TV for direct image reconstruction.	S2S is a recent unsupervised learning method which also uses dropout-based network for denoising an image. We used it as a post-process to denoise the image reconstructed by the FBP method, where the denoising NN is trained using the S2S method with the same configuration as the proposed method.
	The DIP+TV method~\cite{baguer2020computed} combines the DIP approach and TV-based regularization for CT, whose loss function is defined by
	\begin{align}
		\mathcal{L}(\bm{\theta})=	\|\bm{A}f(\bm{z};{\bm{\theta}})-\bm{y}\|_2^2+\alpha \|\nabla f(\bm{z};{\bm{\theta}})\|_1.
	\end{align}
	The DIP+TV is related to the proposed method. Indeed, the proposed method is degenerated to the DIP+TV by setting  set dropout probability $p$ to 1 and using the same input $\bm{z}$.
	In the DIP-based denoising network, a random noise is used as the initial seed. In this paper, the initial seed for the DIP+TV is also set to $\bm{x}_0$, the same as the proposed method.  DIP+TV is trained using the Adam optimizer which the momentum parameter is set as 0.9. The learning rate is set as $10^{-2}$.
	
	In addition, for the dataset of prostate image, we also compared the proposed method with an supervised learning method, FBPConvNet \cite{jin2017deep}.  FBPConvNet is one of a
	representative deep learning methods for CT reconstruction
	that uses the deep NN as a post-processing technique.
	In FBPConvNet, U-net architecture is trained with low -dose and normal-dose image pairs to directly denoise the
	image reconstructed by the FBP method.
	\subsection{Data Simulation}
	\label{CT Reconstruction Problem and Data Simulation}
	We  adopted the proposed method to LDCT reconstruction, in which	$\bm{A}$ is set as the projection matrix.
	By using a monoenergetic source in CT imaging, the  measurements from CT scan follow  Poisson distribution which can be expressed as \cite{ding2016modeling}:
	\begin{eqnarray}
		\bar{\bm{y}}_i\sim {\rm{Poisson}}\{I_i\exp(-[\bm{A}\bm{\bm{x}}]_i)\}+ N(0,\sigma_e^2),
		\label{elemodel}
	\end{eqnarray}
	where $N$ refers to normal distribution, $\bm{x}$ denotes the attenuation map with  $x_j$ being the linear attenuation coefficient in the $j$-th pixel for $j =1,\cdots,n$ and  $n$ denotes the total number of pixels;
	$\bar{\bm{y}}$ represents the measured projection.
	The matrix $\bm{A}$ is the $m \times n$ system matrix with entries $a_{ij}$, and $[\bm{A}\bm{x}]_i=\sum_{j=1}^{n}a_{ij}x_{j}$ denotes the line integral of the attenuation map $\bm{x}$ along the $i$-th X-ray with $i=1,\cdots, m$. $I_i$ is the incident X-ray intensity incorporating X-ray source illumination and the detector efficiency. The noise level is controlled
	by $I_i$, \emph{i.e.}, the noise of measure data
	becomes larger when dose level $I_i$ decreases.  $\sigma_e^2$ denotes the variance of the background electronic noise.	
	To reconstruct the attenuation map $\bm{x}$, we take the logarithm transform on the noisy measurements $\bar{\bm{y}}$ to generate the noisy sinogram $\bm{y}$.

	\subsection{LDCT reconstruction result}
	For quantitative analysis of image quality,  three indices: peak signal to noise ratio (PSNR), root mean square error (RMSE)  and structural similarity index measure (SSIM) \cite{wang2004image} are compared for different reconstruction methods.
	

	\begin{figure}[htbp!]
		\begin{center}		
			\includegraphics[width=1\linewidth,height=0.6\linewidth]{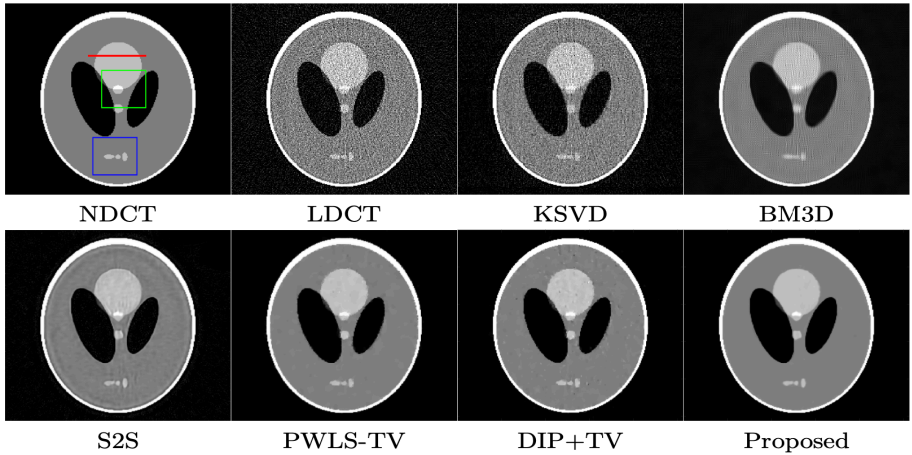}
		\end{center}	
		\caption {Reconstruction results of phantom.}
		\label{Result:phantom}
	\end{figure}

	\begin{figure}[htbp!]
		\begin{center}		
			\includegraphics[width=1\linewidth,height=0.3\linewidth]{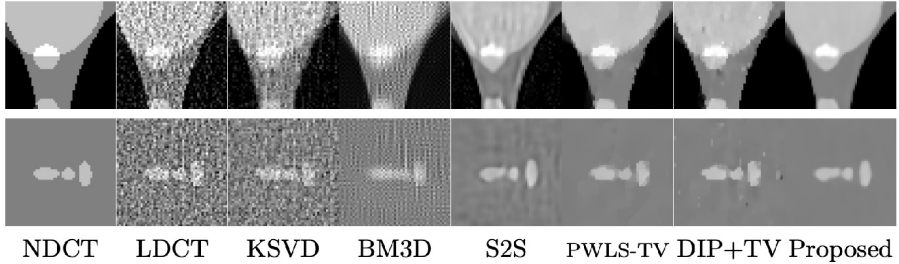}
		\end{center}	
		\caption {Zoomed-in results of Phantom in Fig. \ref{Result:phantom}.}
		\label{Result:phantomZoom}
	\end{figure}

	\begin{table}[htbp!]
		\scalebox{0.8}[0.80]{
			\begin{tabular}{cccccccc}
				\hline
				\hline
				&    LDCT(FBP)&   KSVD&    BM3D&     S2S&          PWLS-TV& DIP+TV& Proposed\\\hline
				PSNR& 25.2472&26.9100&	29.8907& 34.0764&	  35.4817&	  40.5554&	42.3255\\
				RMSE&  0.0273&	0.0226&	 0.0160&  0.0099&	   0.0084&	  0.0047&	0.0038\\
				SSIM&  0.3604&	0.4642&	 0.4453&  0.8343&	   0.9805&	  0.9861&	0.9949\\
				Reconstruction Time&0.01s &145.80s&23.61s&40.07min&3.24s&13.52min&55.22min\\				
				\hline						
				\hline
			\end{tabular}
		}
		\caption{Quantitative reconstruction results of phantom in Fig. \ref{Result:phantom}.}
		\label{SNRRMSE:Phantom}
	\end{table}	
	
		\begin{figure}[ht]
		\begin{center}		
			\subfigure[Profile]{\label{Profile}
				\includegraphics[width=.45\linewidth]{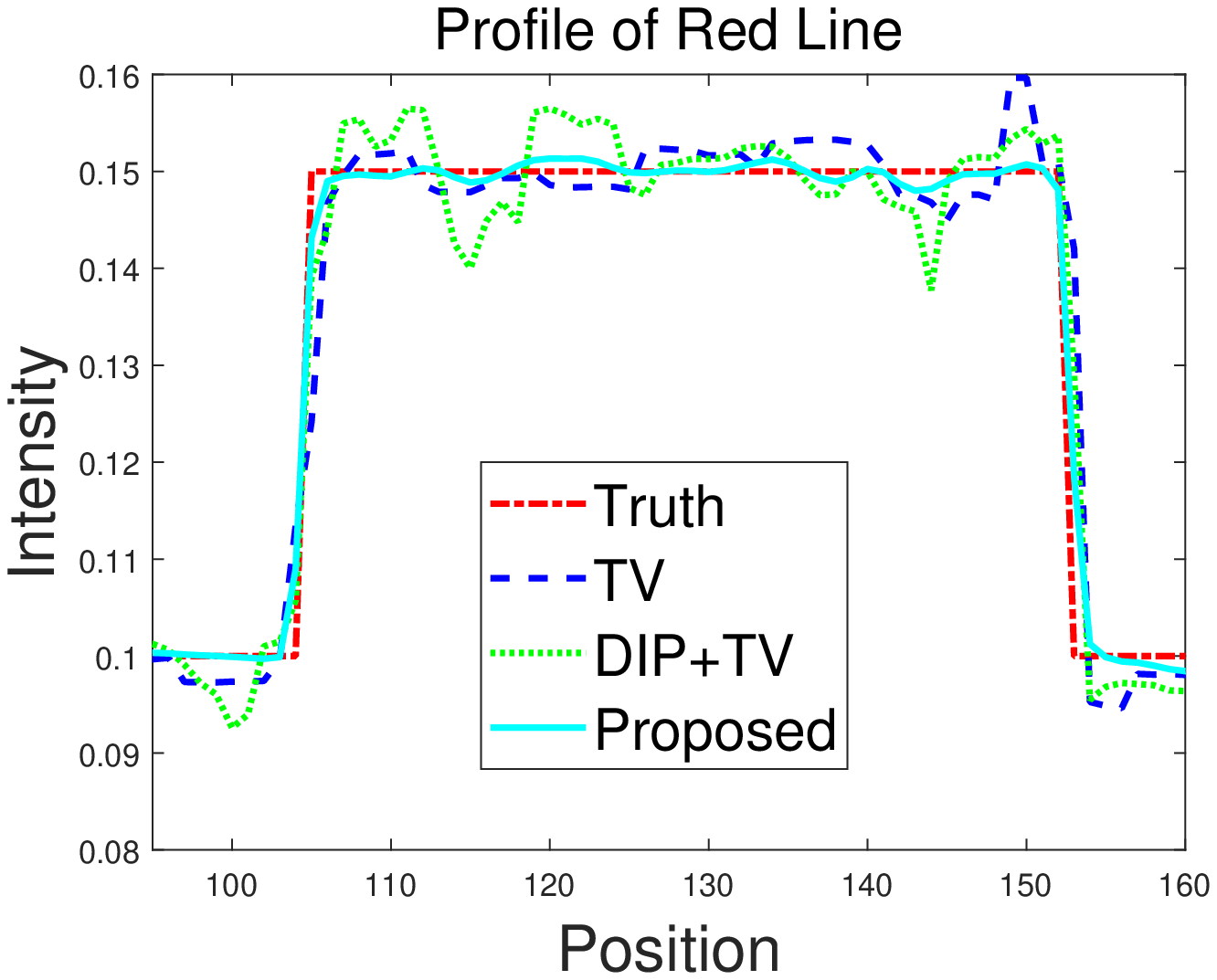}}
			\subfigure[PSNR vs. K]{\label{TestK}
				\includegraphics[width=.45\linewidth]{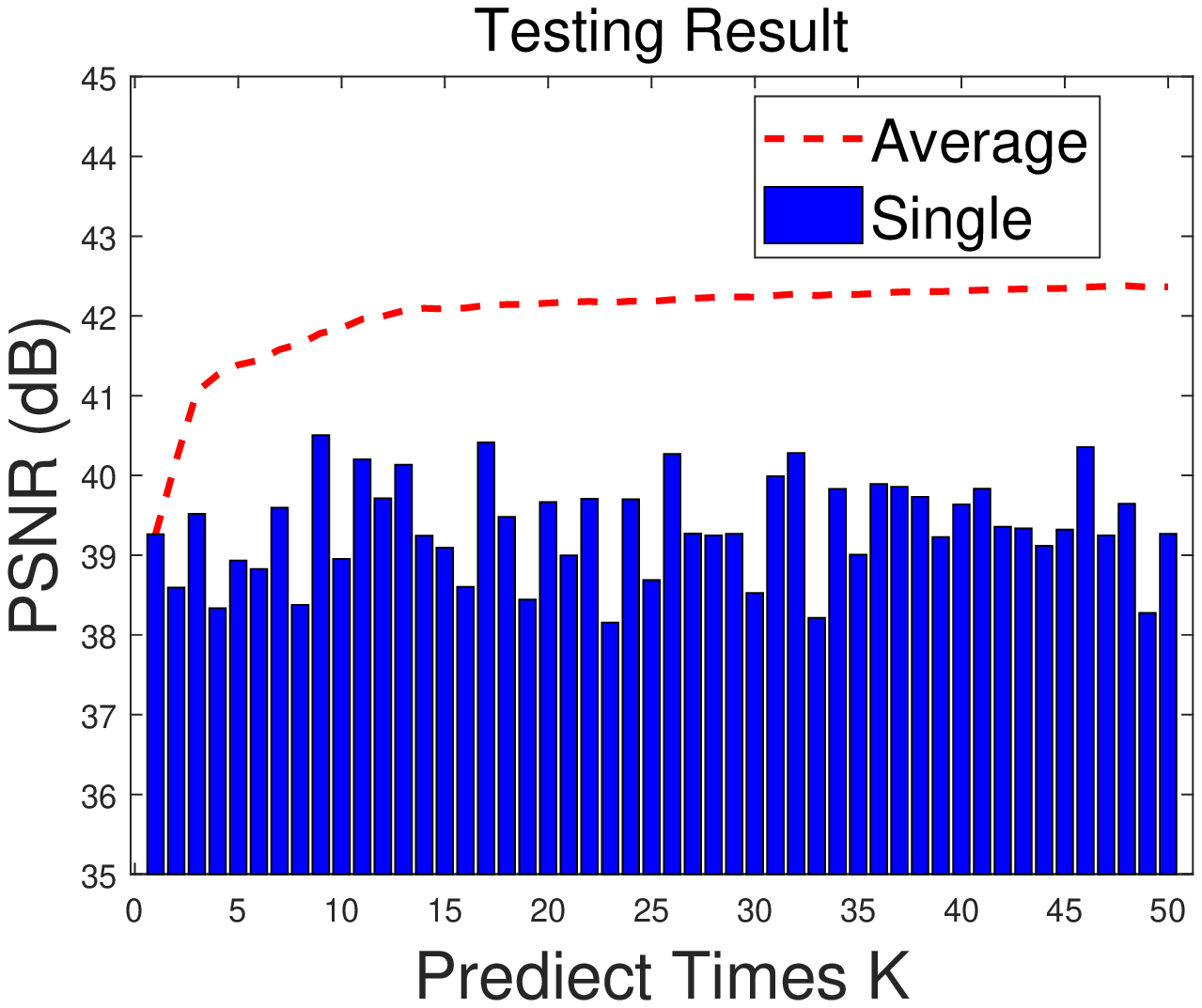}}\\
			\subfigure[Single Prediction 1]{\label{SingleSlice1}
				\includegraphics[width=.3\linewidth,height=.3\linewidth]{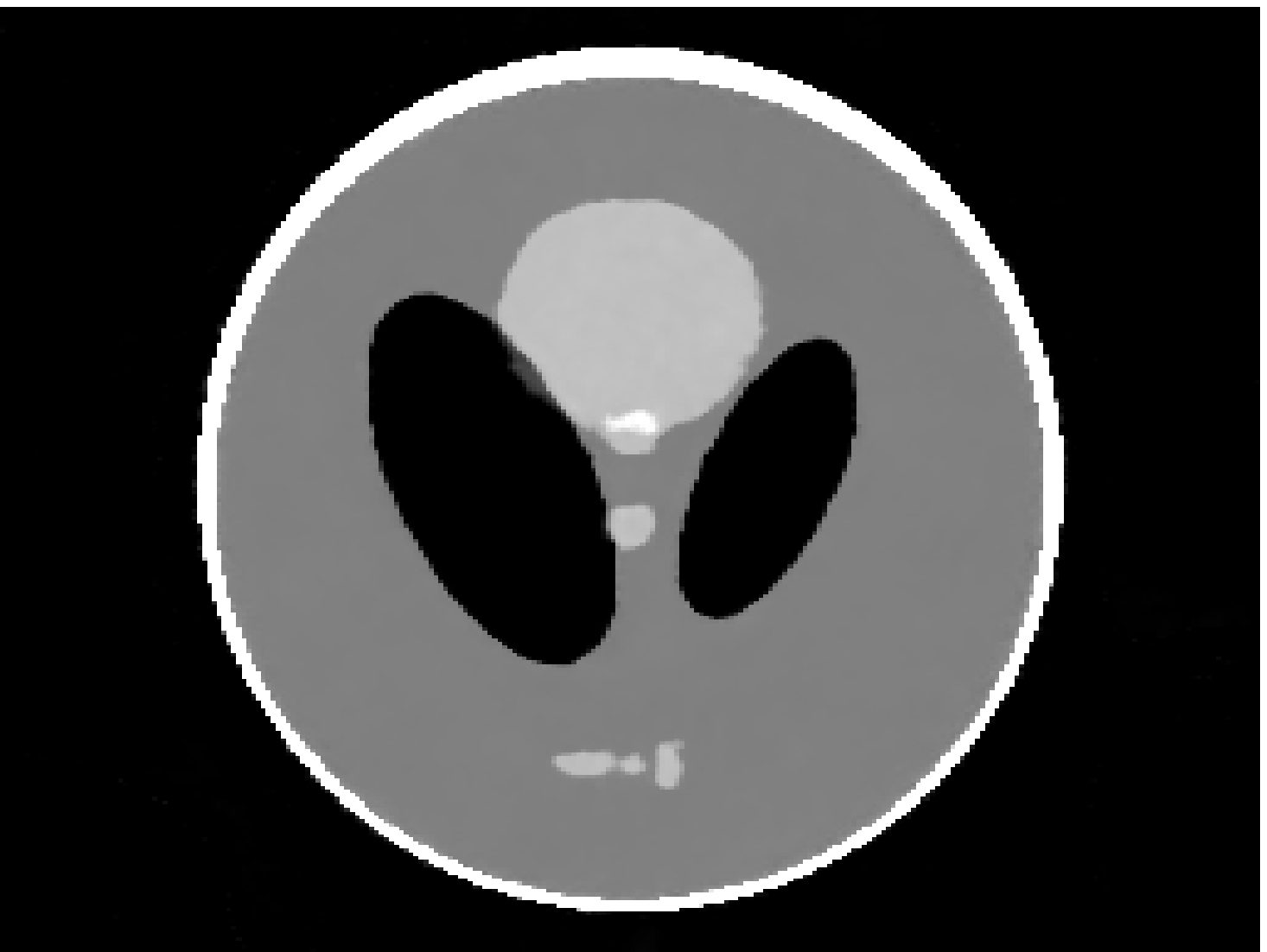}}
			\subfigure[Single Prediction 2]{\label{SingleSlice2}
				\includegraphics[width=.3\linewidth,height=.3\linewidth]{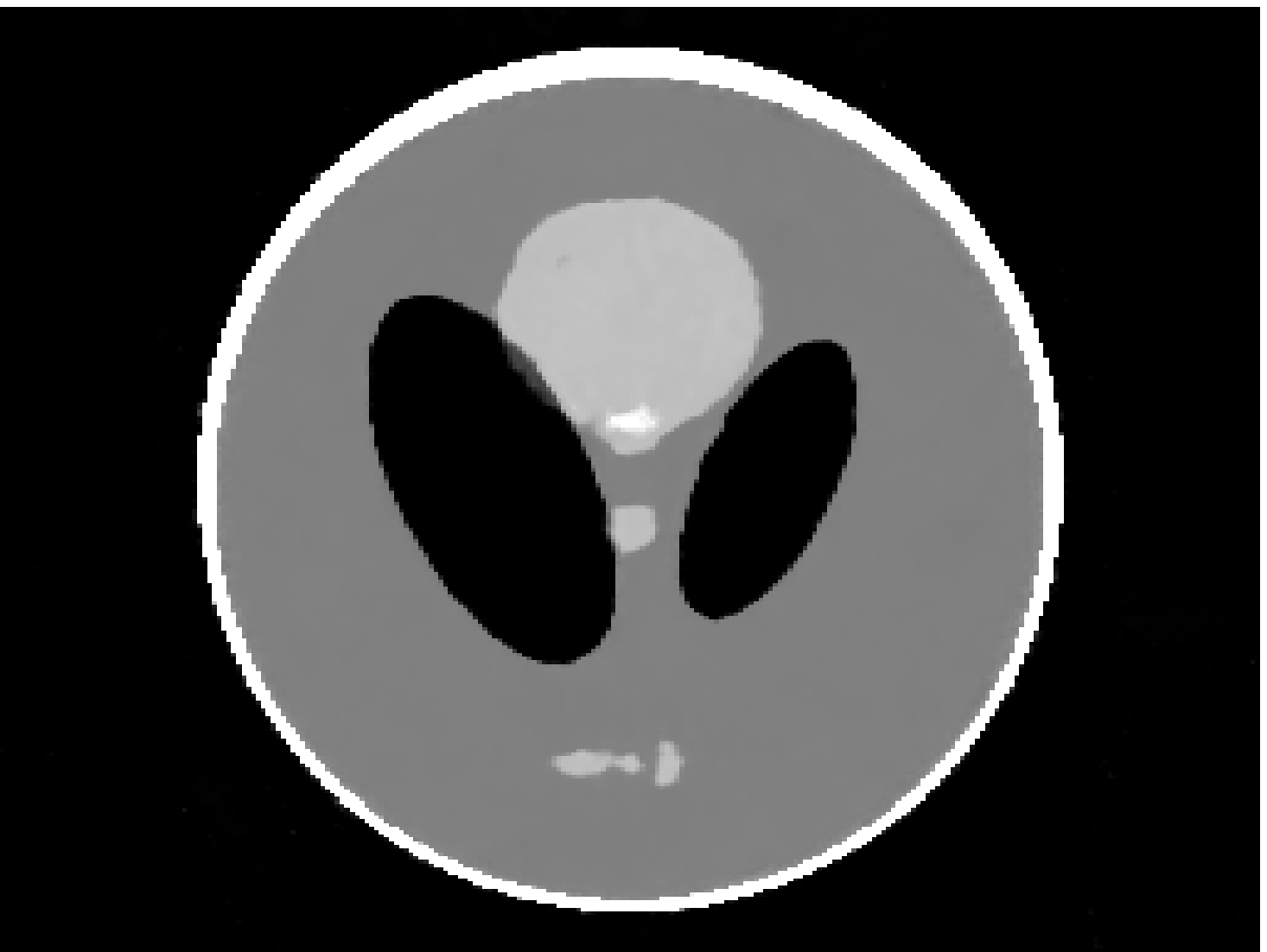}}
			\subfigure[Average]{\label{Average} 
				\includegraphics[width=.3\linewidth,height=.3\linewidth]{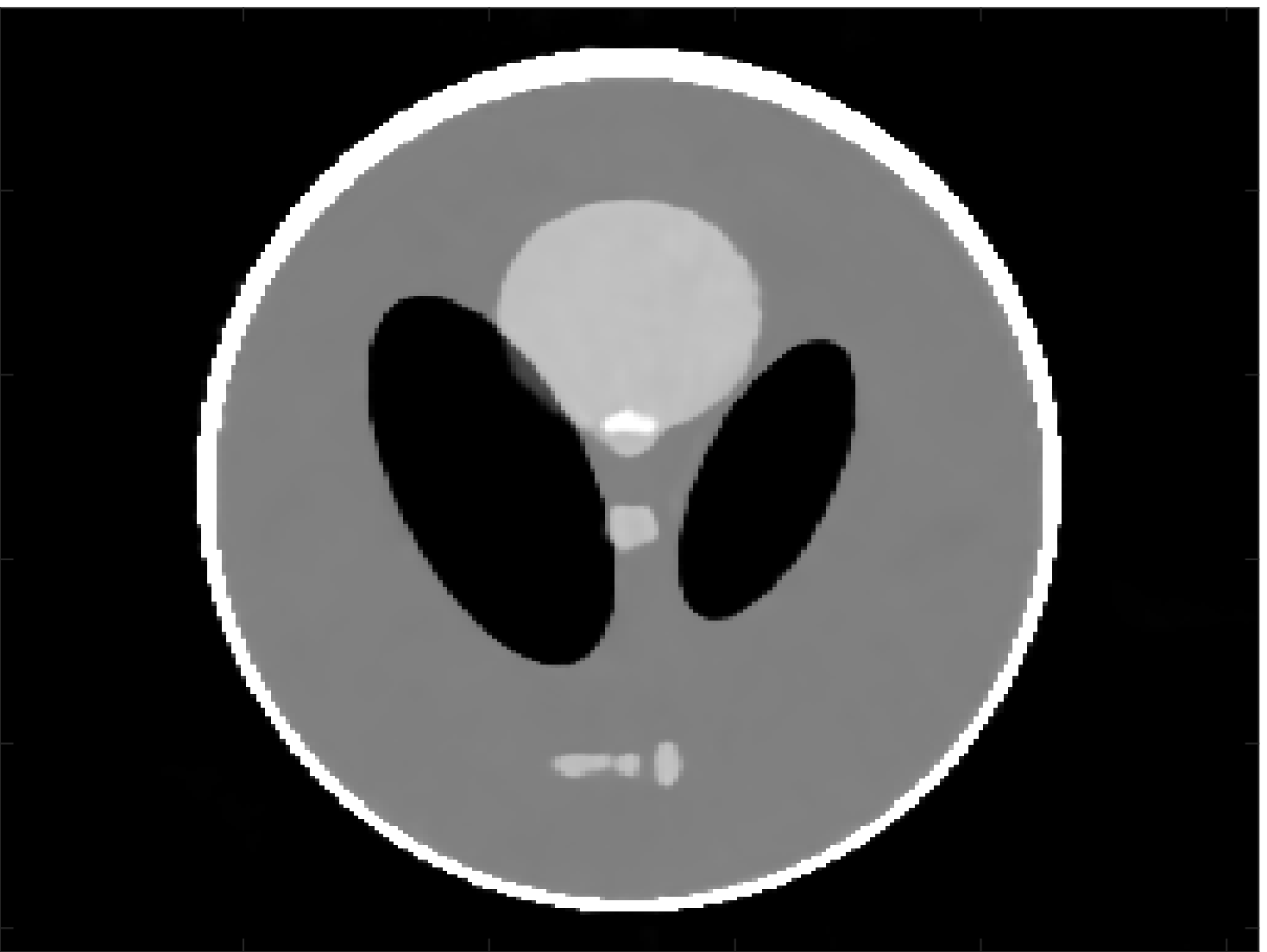}}
		\end{center}
	\caption{Fig.\ref{Profile}: Profile of the line labeled in red  in Fig.\ref{Result:phantom}  NDCT. Fig.\ref{TestK}: PSNR versus prediction times K. Fig.\ref{SingleSlice1} and Fig.\ref{SingleSlice2} are the two single predictions of the image. Fig.\ref{Average}: the average of multiple predictions.}
		\label{K_Profile}
	\end{figure}
	\subsubsection{Phantom image reconstruction}
	To evaluate the effectiveness of the proposed method,
	we simulated the digital phantom of size $256\times256$ and the corresponding  noisy sinogram.
	The LDCT projection data was simulated by adding Poisson noise and the background electronic onto
	the normal-dose projection data  with $I_i=1\times10^3$ and $\sigma_e^2=10$. The simulated geometry for projection data is as follows: fan-beam CT scanner, flat-panel  detector of $0.388~\rm{mm}\times 0.388~\rm{mm}$ pixel size,
	$600$ projection views  evenly  spanning a $360^{\circ}$ circular orbit,  $512$ detector bins for each projection with $1~\rm{mm}$ pixel size, $100.0~\rm{cm}$ source to detector distance and $50.0~\rm{cm}$  source to isocenter distance. The hyper-parameter $\alpha$ was set to 
		$0.02, 0.2$ and  $0.2$ for PWLS-TV, DIP+TV  and the proposed method respectively.

	Fig.\ref{Result:phantom} shows the images reconstructed
	by different methods, and their zoomed-in images of boxes in  Fig. \ref{Result:phantom}
	are displayed in 	Fig. \ref{Result:phantomZoom}.
	With low-dose measurements, LDCT image reconstructed by FBP present large noisy and streaky artifacts.
	In comparison to the zoomed-in NDCT image, the results of
	KSVD, BM3D and S2S  have more streaky artifacts than that of the proposed method.
	In the proposed method, PWLS-TV method and DIP+TV method, TV regularizer can help suppress the noise and remove the artifact in the reconstructed image. 
	Furthermore, the proposed method has a better performance in structure preservation and noise suppression.
	
	Table \ref{SNRRMSE:Phantom} shows quantitative comparison of the results shown in Fig. \ref{Result:phantom}.
	The proposed method has the best performance in terms of three metrics PSNR, RMSE and SSIM.
	Both PWLS-TV and DIP+TV methods improve upon the conventional LDCT result as expected, and the	proposed  outperformed DIP+TV by a noticeable margin, \emph{i.e.} 1.8dB advantage in PSNR.
	Table \ref{SNRRMSE:Phantom} shows the comparison of running time for the proposed method and the other methods. With an NVIDIA A100 graphics card, the reconstruction time of the proposed method is comparable with S2S. For less iterations, DIP+TV method have advantage over the proposed method.

	Fig. \ref{Profile}  shows the profile outlined in red in Fig. \ref{Result:phantom}, where  the results of PWLS-TV, DIP+TV and the proposed method are compared. It can be easily seen that  the cyan line is more close to the ground truth (red line).
	See Fig. \ref{TestK} for the illustration of 
		how the value of $K$, the number of predictions for averaging, impacts the
		performance of the simulated phantom. It shows that the PSNR value steadily increases with the value of $K$ until it hits $15$. Afterward,  the improvement brought by more predictions is rather small. 
		See Fig. \ref{SingleSlice1} and  Fig. \ref{SingleSlice2} for the visualization of two single predictions of the image. It can be seen that there is a  noticeable difference between two single predictions in certain regions. Fig.~\ref{Average} visualizes the average of multiple predictions, which contains fewer artifacts than the two predictions shown in  Fig. \ref{SingleSlice1} and  Fig. \ref{SingleSlice2}. In other words, the average of multiple predictions from the proposed dropout-based method indeed can reduce the artifacts in the reconstructed image.

	\subsubsection{Reconstruction from  different noise level} We evaluated the proposed method with different noise levels.
	For	 patient’s normal dose prostate image with size $256\times256$, we simulated the low-dose
	measurement with dose levels $I_i=1\times10^3$, $5\times10^3$, $1\times10^4$, $5\times10^4$ and $\sigma_e^2=10$. Then, the sinograms of different noise levels were
	obtained by taking logarithm on projection data. The simulated geometry for projection data is the same as that of phantom data simulation.
	
\begin{figure}[htbp!]
		\begin{center}
			\begin{tabular}{c@{\hspace{10pt}}c@{\hspace{10pt}}c@{\hspace{0pt}}c@{\hspace{0pt}}c@{\hspace{0pt}}c@{\hspace{0pt}}c}
				\includegraphics[width=.48\linewidth,height=.32\linewidth]{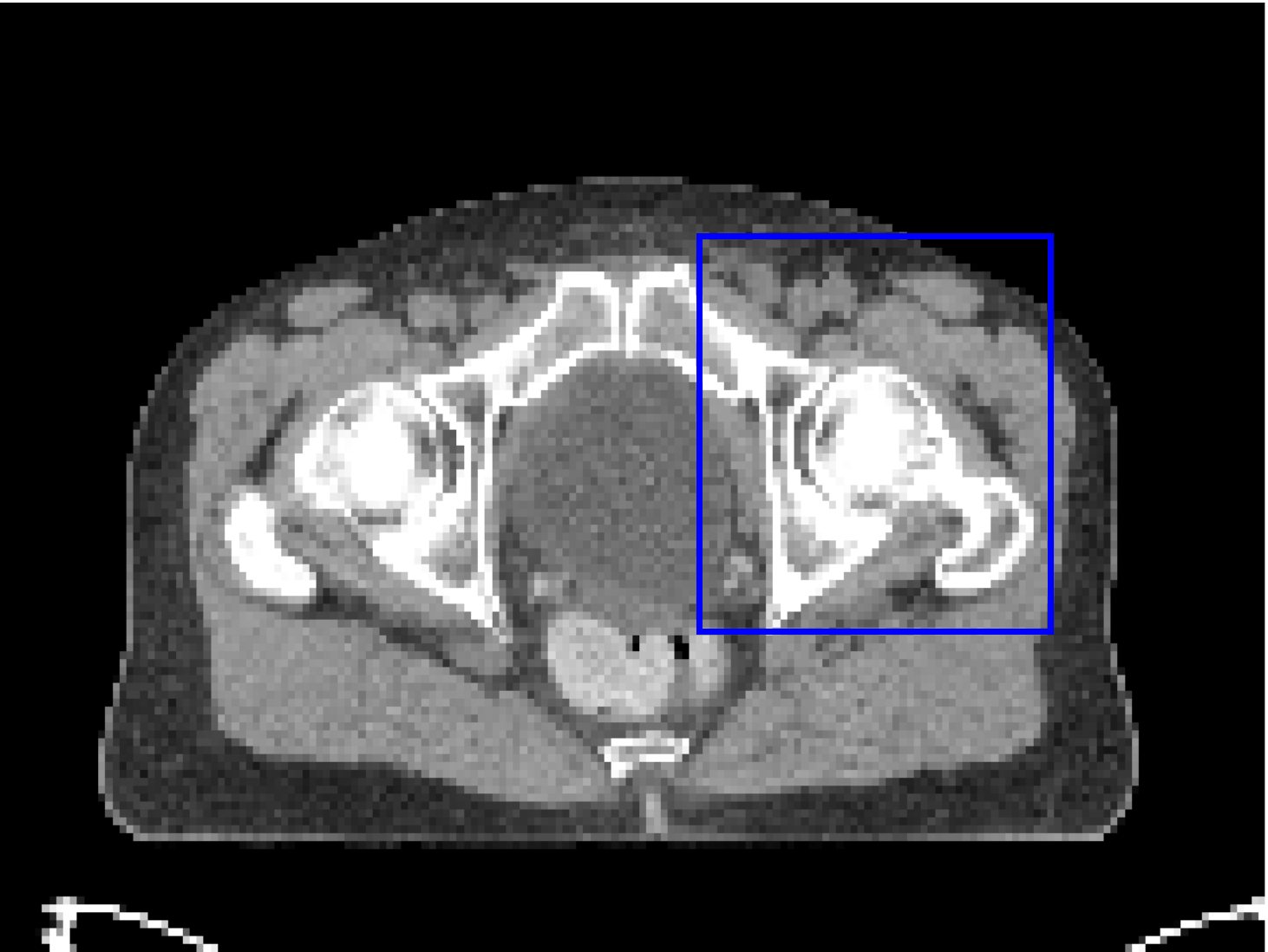}&
				\includegraphics[width=.32\linewidth,height=.32\linewidth]{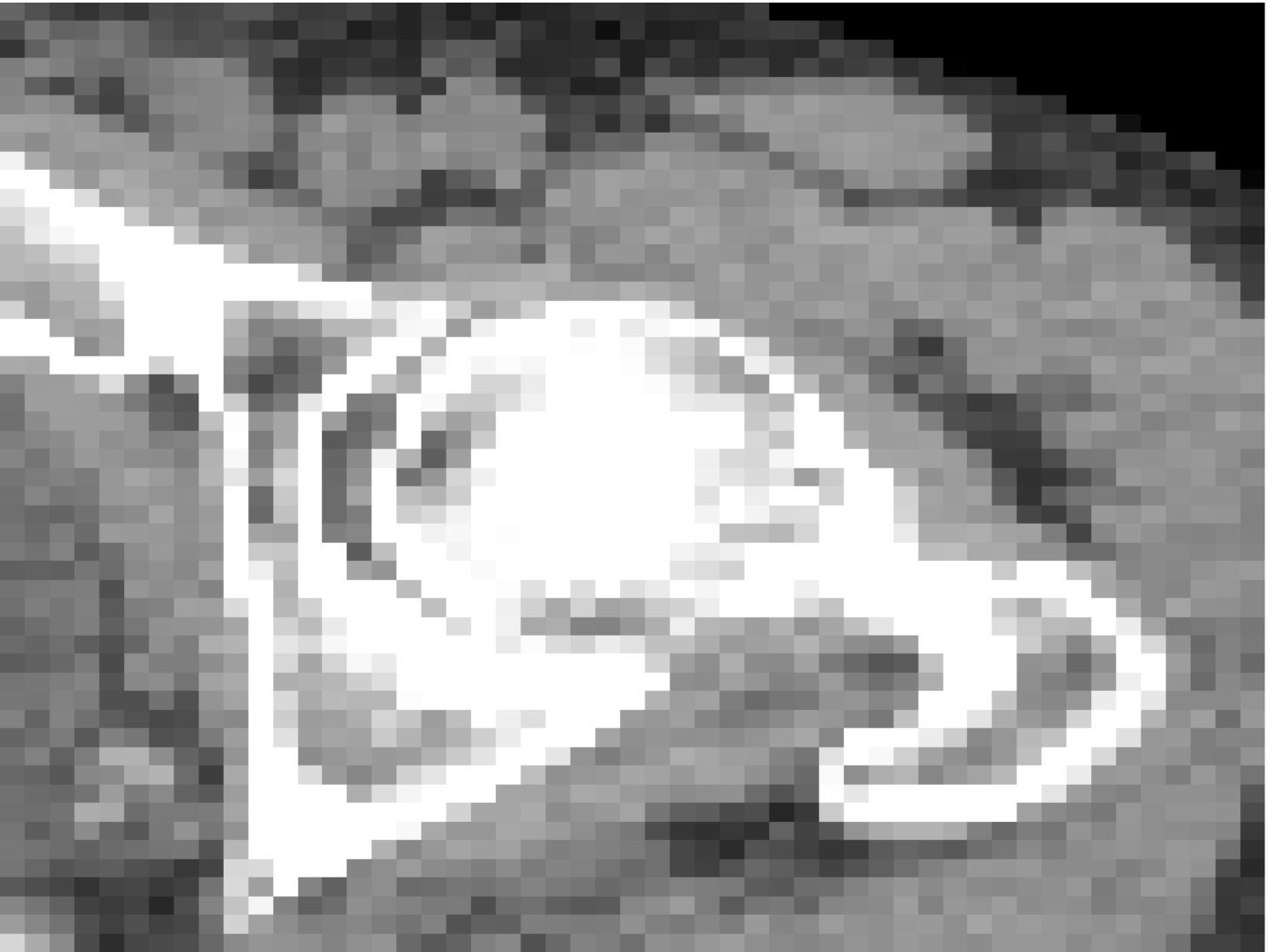}\\
				NDCT&
				ROI
			\end{tabular}
			\caption{Normal dose CT image and a zoomed region (ROI).
			}
			\label{Truth}
		\end{center}
	\end{figure}

\begin{figure*}[htbp!]
		\begin{center}
               \includegraphics[width=1\linewidth,height=1.3\linewidth]{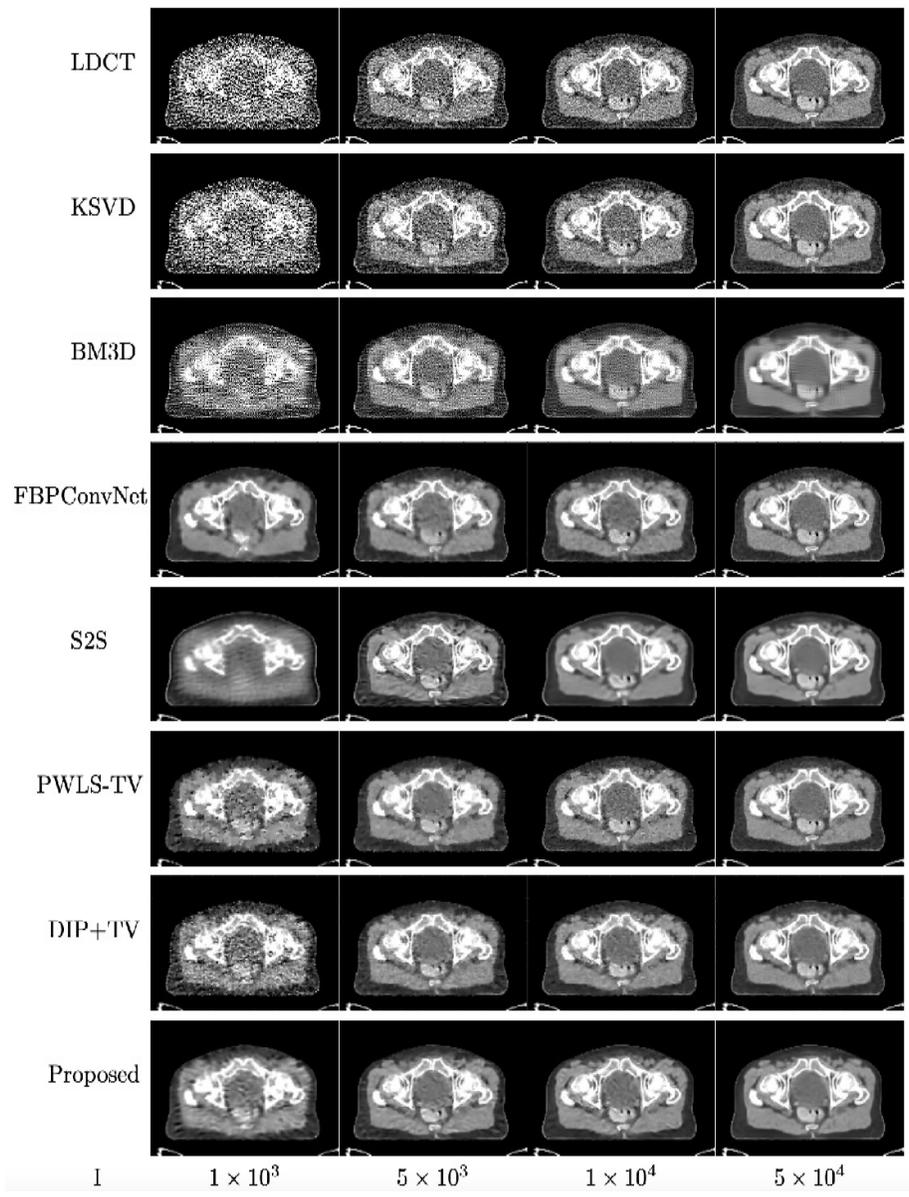}
		\end{center}	
		\caption {Reconstruction results at different dose levels by different methods with $\sigma_e^2=10$}
		\label{Result:prostate}
	\end{figure*}

\begin{figure}[htbp!]
		\begin{center}
                   \includegraphics[width=1\linewidth,height=0.55\linewidth]{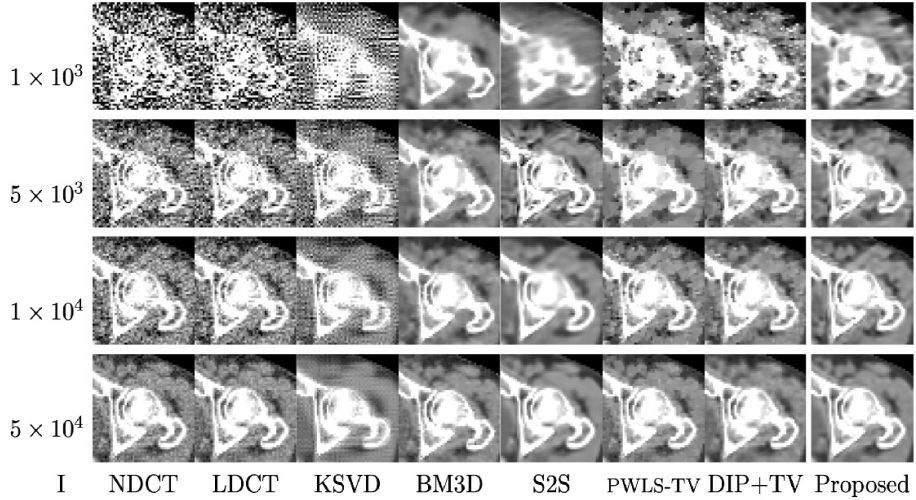}
                   		\end{center}	
		\caption {Zoomed-in results correspond to Fig. \ref{Result:prostate} at different dose levels by different methods with $\sigma_e^2=10$.}
		\label{Result:prostateZoom}
 	\end{figure}

	Fig. \ref{Truth} demonstrates the NDCT
	image and the zoomed region for comparison.
	Fig. \ref{Result:prostate} shows the images reconstructed by
	different methods of different dose levels, and their zoomed-in images of boxes in
	Fig. \ref{Truth} are displayed in Fig. \ref{Result:prostateZoom}. The displayed window is set to $[-150,200]$HU for all figures  with  $\mu_{air} = -1000$HU.
	For all the reconstruction methods, the image quality decreases with the lower dose level. The recovered images by BM3D are not visually satisfactory. The images by  S2S are  blurry that some image details are missing. The proposed method, PWLS-TV method 	and DIP+TV method are the three best performers among all methods without dataset.
	For the three methods with TV regularizer, the proposed method achieved the be  image quality with preserved image structure and less noise.
	In this experiment,  the  values of  $\alpha$ are adjusted to the noise level of the data.
		For PWLS-TV, DIP+TV and the proposed method, $\alpha=0.1,~0.1,~0.1$ with $I=1\times 10^3$; $\alpha=0.05,~0.05,~0.05$ with $I=5\times 10^3$;
		$\alpha=0.02,~0.03,~0.01$ with $I=1\times 10^4$ and  $\alpha=0.01,~0.01,~0.001$ with $I=5\times 10^4$.

	In this prostate dataset, there are 6400 normal-dose prostate
	CT images. We adopted FBPConvNet with  80\% of low-dose and normal dose image pairs. It is shown in Table \ref{SNRRMSEProstate}  and  Fig. \ref{Result:prostate} that supervised method has the best performance in comparison with non-learning methods and unsupervised DL methods. With the dose $I=5\times10^3$ and $I=1\times10^3$, FBPConvNet gained 1-2dB advantage over the proposed method.
	The proposed method and DIP+TV achieved higher PSNR and SSIM, smaller RMSE among all unsupervised methods.
	Moreover, the proposed method  outperformed DIP+TV by 1.0-1.3dB in PSNR.	
	

	\begin{table*}[htbp!]
		\centering	
		\scalebox{0.7}[0.80]{
			\begin{tabular}{cccccccccc}
				\hline
				\hline
				Dose Level&      	      Index& LDCT&      KSVD&       BM3D&  FBPConvNet&     S2S&       PWLS-TV&          DIP+TV& Proposed\\\hline
				\multirow{3}{*}{$1\times10^3$}&PSNR& 16.0859&	16.7163&	19.9995&  29.6054&	25.7518&   25.8050&	    25.3503&	27.5996\\
				&RMSE& 162.8839&	151.4813&	103.8002& 34.3481&	53.5282&   53.2013&	    56.0607&	43.2707\\
				&SSIM& 0.2744&	0.2841&  	0.4379&	   0.8159&  0.5893 &   0.7001 &	     0.6739&	0.7675  \\\hline
				\multirow{3}{*}{$5\times10^3$}&PSNR& 24.3676&	24.9922&	26.2975&  33.2609&	30.3793&   30.0352&	    30.8064&	31.9237\\
				&RMSE& 62.7758&	58.4203&	50.2687&  26.3339&	31.4200&   32.6897&	    29.9125&	26.3018\\
				&SSIM& 0.5510&	0.5805&	    0.6622&	   0.8762&  0.8086 &   0.8535 &	     0.8544&	0.8722 \\\hline
				\multirow{3}{*}{$1\times10^4$}&PSNR& 27.4944&	27.5416&	29.1881&  33.3785&	31.3774&   31.3655&     32.4104&	33.5989\\
				&RMSE& 43.7979&	43.5605&	36.0386&  22.2457&	28.0094&   28.0477&	    24.8687&	21.6883\\
				&SSIM& 0.6834&	0.7024&  	0.7808&	   0.8943&  0.8702&   0.8612&  	 0.8894&	0.9070\\\hline
				\multirow{3}{*}{$5\times10^4$}&PSNR& 33.1157&	32.5643&	33.6708&   36.9016& 34.3365&   35.0872&	    35.4009&	36.1700\\
				&RMSE& 22.9291&	24.4320&	21.5097&   14.8283&	19.9226&   18.2731&	    17.3623&	16.1314\\
				&SSIM& 0.8861&	0.8894&	    0.9166&    0.9518&	0.9301&   0.9234&	    0.9220&	    0.9408\\
				\hline						
				\hline
			\end{tabular}
		}
		\caption{Quantitative reconstruction results  of Fig. \ref{Result:prostate}.}		
		\label{SNRRMSEProstate}	
	\end{table*}
	\subsubsection{Clinical data reconstruction}
	\begin{figure}[ht]
		\begin{center}		
			\subfigure[]{\label{Result:mayo20_50000}
				\includegraphics[width=.4\linewidth,height=.4\linewidth]{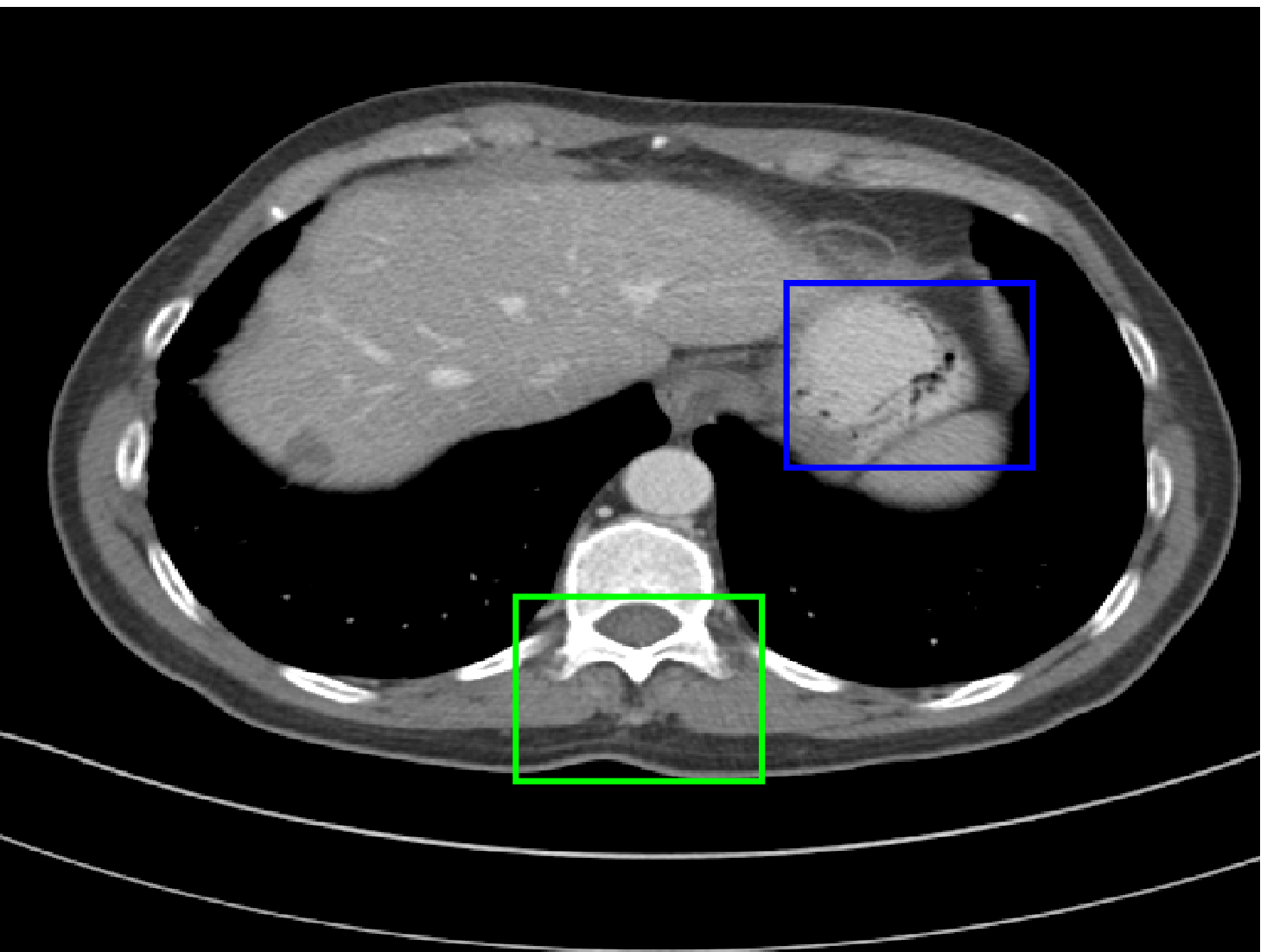}}
			\subfigure[]{\label{Result:mayo35_50000}
				\includegraphics[width=.4\linewidth,height=.4\linewidth]{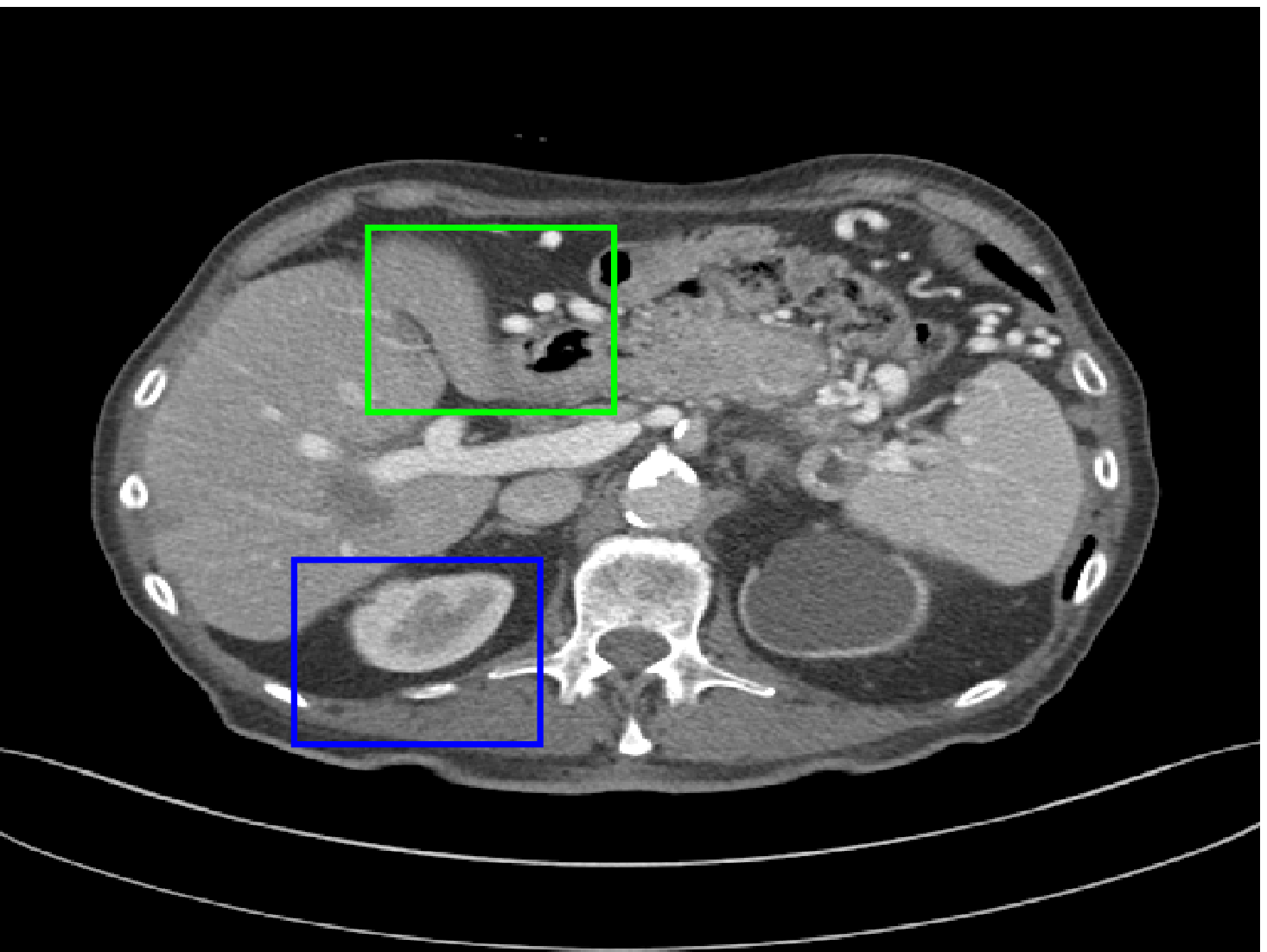}}
		\end{center}
		\caption{Two normal dose CT images from mayo dataset. }
		\label{mayo_image}
	\end{figure}
	To evaluate the performance of the proposed method 	under realistic conditions,  clinical image was
	used, which was established  by Mayo Clinics for “the 2016 NIH-AAPM-Mayo Clinic Low Dose CT Grand
	Challenge”.
	We aimed to reconstruct  the image with size $512\times 512$ from the projection data of
	full dose CT images of 3mm thickness.
	The projection data  is composed of projection data from $600$ projection views  evenly  spanning a $360^{\circ}$ circular orbit,  $768$ detector bins for each projection, $100.0~\rm{cm}$ source to detector distance and $50.0~\rm{cm}$  source to isocenter distance.
	
\begin{figure}[htbp!]
		\begin{center}			
				\includegraphics[width=1\linewidth,height=.3\linewidth]{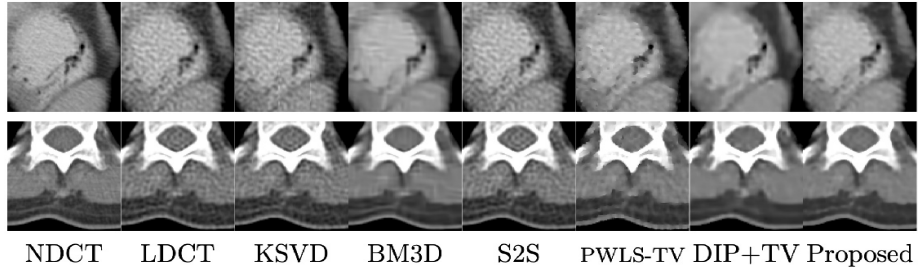}	
		\end{center}	
		\caption {Zoomed-in results of Mayo image correspond to Fig. \ref{Result:mayo20_50000} .}
		\label{Result:mayo20_50000Zoom}
	\end{figure}

\begin{figure}[htbp!]
		\begin{center}		
				\includegraphics[width=1\linewidth,height=.3\linewidth]{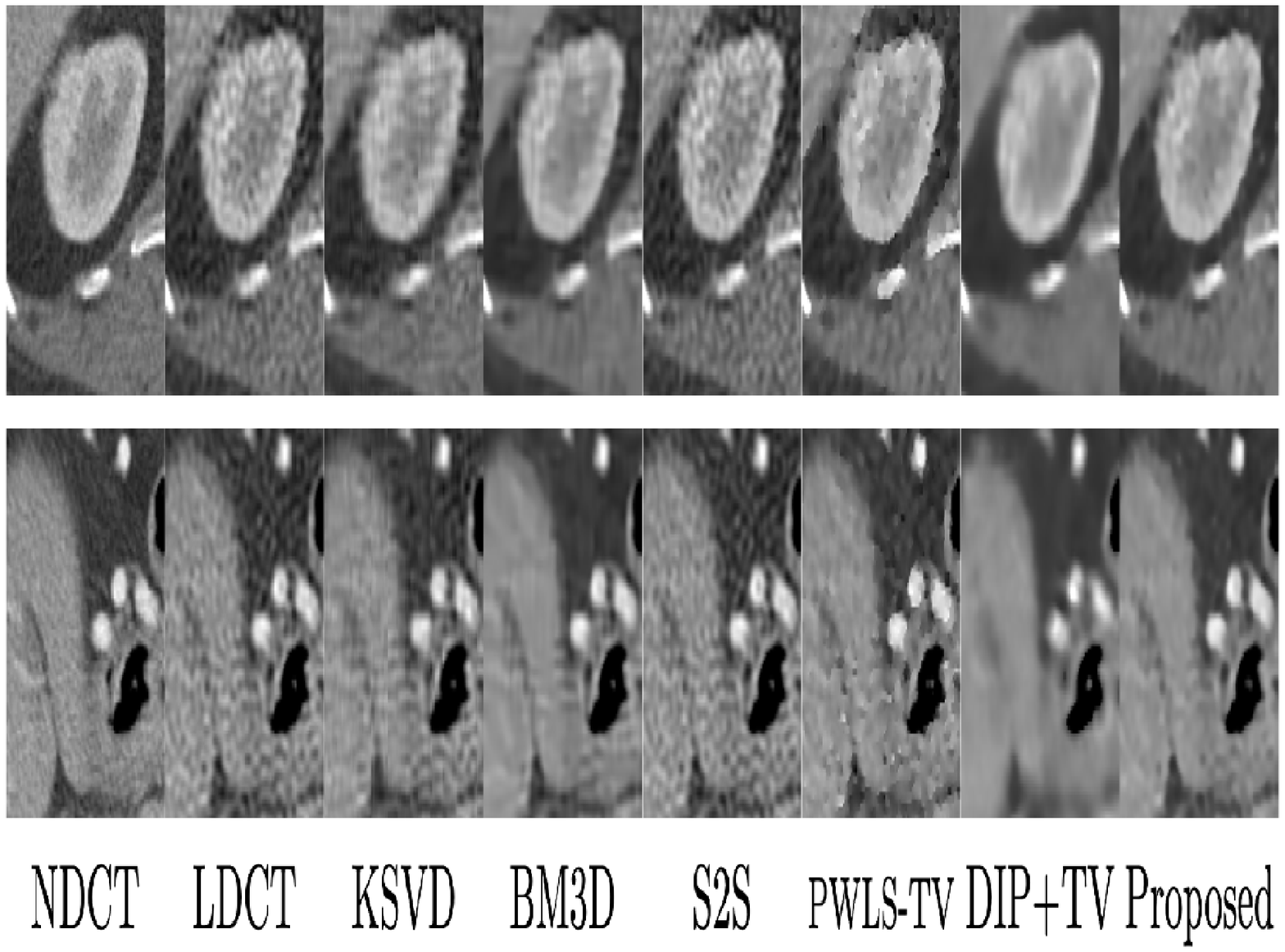}
		\end{center}	
		\caption {Zoomed-in results of Mayo image correspond to Fig. \ref{Result:mayo35_50000} .}
		\label{Result:mayo35_50000Zoom}
	\end{figure}

	Fig.~\ref{mayo_image} shows two normal dose mayo image slices and two ROIs of each slice are labeled by blue and green boxes.
	Fig.~\ref{Result:mayo20_50000Zoom} and Fig.~\ref{Result:mayo35_50000Zoom} demonstrate the zoomed region of interest correspond to Fig.~\ref{Result:mayo20_50000} and Fig.~\ref{Result:mayo35_50000} respectively.
	The displayed window is set to $[-200,400]$HU for all figures  with $\mu_{air} = -1000$HU. LDCT image is reconstructed
	by FBP with heavy noise and artifacts. Both post-processing  type methods BM3D and S2S are not able to remove the streaky artifacts. 
	TV and DIP+TV could reduce noise and remove artifacts, but the image details are smoothed out.
	The proposed method achieves the image result with better  structure preservation and noise suppression.
	For the image in Fig.~\ref{Result:mayo20_50000}, the hyper-parameter $\alpha$ was set to 0.5, 0.3 and 0.3 for PWLS-TV, DIP+TV and the proposed method respectively for optimal performance.
		For the image in Fig.~\ref{Result:mayo35_50000}, the hyper-parameter $\alpha$ was set to 0.3 for all the TV-based methods.
	
	Quantitative reconstruction results corresponding to the image in Fig.~\ref{Result:mayo20_50000} and Fig.~\ref{Result:mayo35_50000}
	are given in Table \ref{MayoSNRResult}. The proposed method has the best performance in terms of the metrics than the other methods.
	In comparison with DIP+TV, the proposed method outperforms DIP+TV in PSNR about 1dB.
	
	\begin{table}[htbp!]
		\centering	
		\scalebox{0.85}[0.85]{
			\begin{tabular}{cccccccccc}
				\hline
				\hline
				&	Index	                                              & LDCT   &KSVD &      BM3D &     S2S&      \footnotesize{PWLS-TV}&     DIP+TV& Proposed\\\hline
				\multirow{3}{*}{Fig. \ref{Result:mayo20_50000}}         &PSNR& 30.0174&	29.9941&	30.4400& 30.1597& 34.3136&	    34.7112&   36.4061\\
				&RMSE& 42.4475&	42.5617&	40.4317& 41.7577& 25.8847&   	24.7265&   20.3432\\
				&SSIM& 0.9098 &	0.9113 &  	0.9398 &  0.9138&  0.9475&	     0.9518&	0.9577  \\\hline
				\multirow{3}{*}{Fig. \ref{Result:mayo35_50000}}         &PSNR& 31.011 &	31.0529&	31.6613& 31.3406& 33.9875&	   33.7172&	35.2124\\
				&RMSE& 41.9409&	41.7388&	38.9155& 40.3789& 29.7721&	    30.7134&	25.8562\\
				&SSIM& 0.8921 &	0.8950 &	 0.9273&  0.9095&  0.9276&	    0.9192&   0.9461 \\\hline
				\hline						
			\end{tabular}
		}	
		\caption{Quantitative reconstruction results of Fig. \ref{Result:mayo20_50000} and Fig. \ref{Result:mayo35_50000}. }
		\label{MayoSNRResult}
	\end{table}	
	\section{Discussion and conclusion}
	\label{conclusion}		
	In this paper, we proposed an unsupervised learning method for LDCT image reconstruction, which is based on a re-parametrization via the network with random weights for Bayesian inference. The proposed method does not require any external training samples, which is flexible and easy to use in practice. The experiments showed that the proposed method out- performed the representative non-learning-based methods and two recent unsupervised DL methods. The proposed method can be potentially adapted to other medical image reconstruc- tion problems, where the training samples are costly or difficult to collect, \emph{e.g.}  sparse-view CT reconstruction and image reconstruction from sparse samples in MRI. In the future, we would like investigate better data-driven regularization for avoiding the possible over-fitting caused by the absence of ground truth images. In this paper, the values of the regularization parameter and dropout probability are manually tuned up for optimal performance. How to automate the setting of these parameters remains a question to be studied in future work. In addition, unsupervised DL methods cannot pre-train a model such that it can be called to process test image without training. Computational efficiency is important for practical usage. It will be our future work on how to address such an issue. One direction is to study a light-weight network with few parameters for LDCT. Another direction is to study test-time adaption which uses an unsupervised method to quickly adapt a pre-trained model to process test data.

	\section*{Acknowledgement}
Qiaoqiao Ding and Xiaoqun Zhang would like to acknowledge the support from National Natural Science Foundation of China (Grant No. 12090024) and the Sino-German Mobil- ity Programme (M-0187) from Sino-German Center for Research Promotion. Yuhui Quan would like to acknowledge the support from National Natural Science Foundation of China (Grant No. 61872151) and Natural Science Foundation of Guangdong Province (Grant No. 2022A1515011755). Hui Ji would like to acknowledge the support from the Singapore MOE AcRF research grant (R146000229114). We thank the Student Innovation Center at Shanghai Jiao Tong University for providing us the computing services.
	\bibliographystyle{plain}
	\bibliography{biblist.bib}

\begin{thebibliography}{10}

\bibitem{adler2018learned}
Jonas Adler and Ozan {\"O}ktem.
\newblock Learned primal-dual reconstruction.
\newblock {\em IEEE transactions on medical imaging}, 37(6):1322--1332, 2018.

\bibitem{baguer2020computed}
Daniel~Otero Baguer, Johannes Leuschner, and Maximilian Schmidt.
\newblock Computed tomography reconstruction using deep image prior and learned
  reconstruction methods.
\newblock {\em Inverse Problems}, 36(9):094004, 2020.

\bibitem{balda2012ray}
Michael Balda, Joachim Hornegger, and Bjoern Heismann.
\newblock Ray contribution masks for structure adaptive sinogram filtering.
\newblock {\em IEEE transactions on medical imaging}, 31(6):1228--1239, 2012.

\bibitem{batson2019noise2self}
Joshua Batson and Loic Royer.
\newblock {Noise2Self: Blind Denoising by Self-Supervision}.
\newblock In {\em International Conference on Machine Learning}, pages
  524--533, 2019.

\bibitem{cai2014cine}
Jianfeng Cai, Xun Jia, Hao Gao, Steve~B Jiang, Zuowei Shen, and Hongkai Zhao.
\newblock {Cine cone beam CT reconstruction using low-rank matrix
  factorization: algorithm and a proof-of-principle study}.
\newblock {\em IEEE transactions on medical imaging}, 33(8):1581--1591, 2014.

\bibitem{chen2017low}
Hu~Chen, Yi~Zhang, Mannudeep~K Kalra, Feng Lin, Yang Chen, Peixi Liao, Jiliu
  Zhou, and Ge~Wang.
\newblock Low-dose {CT} with a residual encoder-decoder convolutional neural
  network.
\newblock {\em IEEE transactions on medical imaging}, 36(12):2524--2535, 2017.

\bibitem{chen2017lowS}
Hu~Chen, Yi~Zhang, Weihua Zhang, Peixi Liao, Ke~Li, Jiliu Zhou, and Ge~Wang.
\newblock Low-dose {CT} via convolutional neural network.
\newblock {\em Biomedical optics express}, 8(2):679--694, 2017.

\bibitem{chen2022nonblind}
Mingqin Chen, Yuhui Quan, Tongyao Pang, and Hui Ji.
\newblock Nonblind image deconvolution via leveraging model uncertainty in an
  untrained deep neural network.
\newblock {\em International Journal of Computer Vision}, pages 1--20, 2022.

\bibitem{dabov2007image}
Kostadin Dabov, Alessandro Foi, Vladimir Katkovnik, and Karen Egiazarian.
\newblock Image denoising by sparse 3-d transform-domain collaborative
  filtering.
\newblock {\em IEEE Transactions on image processing}, 16(8):2080--2095, 2007.

\bibitem{ding2020low}
Qiaoqiao Ding, Gaoyu Chen, Xiaoqun Zhang, Qiu Huang, Hui Ji, and Hao Gao.
\newblock {Low-dose CT with deep learning regularization via proximal
  forward--backward splitting}.
\newblock {\em Physics in Medicine \& Biology}, 65(12):125009, 2020.

\bibitem{ding2021learnable}
Qiaoqiao Ding, Hui Ji, Hao Gao, and Xiaoqun Zhang.
\newblock Learnable multi-scale fourier interpolation for sparse view ct image
  reconstruction.
\newblock In {\em International Conference on Medical Image Computing and
  Computer-Assisted Intervention}, pages 286--295. Springer, 2021.

\bibitem{ding2016modeling}
Qiaoqiao Ding, Yong Long, Xiaoqun Zhang, and Jeffrey~A Fessler.
\newblock Modeling mixed {Poisson-Gaussian} noise in statistical image
  reconstruction for {X-ray} {CT}.
\newblock {\em Proc. 4th Intl. Mtg. on image formation in X-ray CT}, pages
  {399--402}, 2016.

\bibitem{ding2021deep}
Qiaoqiao Ding, Yuesong Nan, Hao Gao, and Hui Ji.
\newblock {Deep Learning with Adaptive Hyper-parameters for Low-Dose CT Image
  Reconstruction}.
\newblock {\em IEEE Transactions on Computational Imaging}, 7:648--660, 2021.

\bibitem{feldkamp1984practical}
Lee~A Feldkamp, Lloyd~C Davis, and James~W Kress.
\newblock Practical cone-beam algorithm.
\newblock {\em Josa a}, 1(6):612--619, 1984.

\bibitem{gal2016dropout}
Yarin Gal and Zoubin Ghahramani.
\newblock Dropout as a bayesian approximation: Representing model uncertainty
  in deep learning.
\newblock In {\em international conference on machine learning}, pages
  1050--1059. PMLR, 2016.

\bibitem{gong2018pet}
Kuang Gong, Ciprian Catana, Jinyi Qi, and Quanzheng Li.
\newblock {PET image reconstruction using deep image prior}.
\newblock {\em IEEE transactions on medical imaging}, 38(7):1655--1665, 2018.

\bibitem{gupta2018cnn}
Harshit Gupta, Kyong~Hwan Jin, Ha~Q Nguyen, Michael~T McCann, and Michael
  Unser.
\newblock {CNN}-based projected gradient descent for consistent {CT} image
  reconstruction.
\newblock {\em IEEE transactions on medical imaging}, 37(6):1440--1453, 2018.

\bibitem{hasan2020hybrid}
Ahmed~M Hasan, Mohammad~Reza Mohebbian, Khan~A Wahid, and Paul Babyn.
\newblock {Hybrid Collaborative Noise2Noise Denoiser for Low-dose CT Images}.
\newblock {\em IEEE Transactions on Radiation and Plasma Medical Sciences},
  2020.

\bibitem{he2018optimizing}
Ji~He, Yan Yang, Yongbo Wang, Dong Zeng, Zhaoying Bian, Hao Zhang, Jian Sun,
  Zongben Xu, and Jianhua Ma.
\newblock {Optimizing a parameterized plug-and-play ADMM for iterative low-dose
  CT reconstruction}.
\newblock {\em IEEE transactions on medical imaging}, 38(2):371--382, 2018.

\bibitem{he2015delving}
Kaiming He, Xiangyu Zhang, Shaoqing Ren, and Jian Sun.
\newblock Delving deep into rectifiers: Surpassing human-level performance on
  imagenet classification.
\newblock In {\em Proceedings of the IEEE international conference on computer
  vision}, pages 1026--1034, 2015.

\bibitem{hendriksen2020noise2inverse}
Allard~Adriaan Hendriksen, Dani{\"e}l~Maria Pelt, and K~Joost Batenburg.
\newblock {Noise2Inverse: Self-Supervised Deep Convolutional Denoising for
  Tomography}.
\newblock {\em IEEE Transactions on Computational Imaging}, 6:1320--1335, 2020.

\bibitem{hsieh2003computed}
Jiang Hsieh.
\newblock {\em Computed tomography: principles, design, artifacts, and recent
  advances}, volume 114.
\newblock SPIE press, 2003.

\bibitem{jia2011gpu}
Xun Jia, Bin Dong, Yifei Lou, and Steve~B Jiang.
\newblock {GPU}-based iterative cone-beam {CT} reconstruction using tight frame
  regularization.
\newblock {\em Physics in Medicine \& Biology}, 56(13):3787, 2011.

\bibitem{jia20104d}
Xun Jia, Yifei Lou, Bin Dong, Zhen Tian, and Steve Jiang.
\newblock {4D computed tomography reconstruction from few-projection data via
  temporal non-local regularization}.
\newblock In {\em International Conference on Medical Image Computing and
  Computer-Assisted Intervention}, pages 143--150. Springer, 2010.

\bibitem{jin2017deep}
Kyong~Hwan Jin, Michael~T McCann, Emmanuel Froustey, and Michael Unser.
\newblock Deep convolutional neural network for inverse problems in imaging.
\newblock {\em IEEE Transactions on Image Processing}, 26(9):4509--4522, 2017.

\bibitem{kak2002principles}
Avinash~C Kak, Malcolm Slaney, and Ge~Wang.
\newblock Principles of computerized tomographic imaging, 2002.

\bibitem{lehtinen2018noise2noise}
Jaakko Lehtinen, Jacob Munkberg, Jon Hasselgren, Samuli Laine, Tero Karras,
  Miika Aittala, and Timo Aila.
\newblock {Noise2Noise: Learning Image Restoration without Clean Data}.
\newblock In {\em ICML}, 2018.

\bibitem{li2017low}
Heyi Li and Klaus Mueller.
\newblock Low-dose {CT} streak artifacts removal using deep residual neural
  network.
\newblock In {\em Proc. Fully Three-Dimensional Image Reconstruction Radiol.
  Nucl. Med.(Fully3D)}, pages 191--194, 2017.

\bibitem{li2022supervised}
Ji~Li, Yuesong Nan, and Hui Ji.
\newblock Un-supervised learning for blind image deconvolution via monte-carlo
  sampling.
\newblock {\em Inverse Problems}, 38(3):035012, 2022.

\bibitem{manduca2009projection}
Armando Manduca, Lifeng Yu, Joshua~D Trzasko, Natalia Khaylova, James~M Kofler,
  Cynthia~M McCollough, and Joel~G Fletcher.
\newblock Projection space denoising with bilateral filtering and {CT} noise
  modeling for dose reduction in {CT}.
\newblock {\em Medical physics}, 36(11):4911--4919, 2009.

\bibitem{paszke2017automatic}
Adam Paszke, Sam Gross, Soumith Chintala, Gregory Chanan, Edward Yang, Zachary
  DeVito, Zeming Lin, Alban Desmaison, Luca Antiga, and Adam Lerer.
\newblock Automatic differentiation in pytorch.
\newblock {\em Neural Information Processing Systems}, 2017.

\bibitem{quan2020self2self}
Yuhui Quan, Mingqin Chen, Tongyao Pang, and Hui Ji.
\newblock {Self2Self With Dropout: Learning Self-Supervised Denoising From
  Single Image}.
\newblock In {\em Proceedings of the IEEE/CVF Conference on Computer Vision and
  Pattern Recognition}, pages 1890--1898, 2020.

\bibitem{sidky2006accurate}
Emil~Y Sidky, Chien-Min Kao, and Xiaochuan Pan.
\newblock Accurate image reconstruction from few-views and limited-angle data
  in divergent-beam {CT}.
\newblock {\em Journal of X-ray Science and Technology}, 14(2):119--139, 2006.

\bibitem{sidky2008image}
Emil~Y Sidky and Xiaochuan Pan.
\newblock Image reconstruction in circular cone-beam computed tomography by
  constrained, total-variation minimization.
\newblock {\em Physics in Medicine \& Biology}, 53(17):4777, 2008.

\bibitem{ulyanov2018deep}
Dmitry Ulyanov, Andrea Vedaldi, and Victor Lempitsky.
\newblock Deep image prior.
\newblock In {\em Proceedings of the IEEE Conference on Computer Vision and
  Pattern Recognition}, pages 9446--9454, 2018.

\bibitem{van2018compressed}
Dave Van~Veen, Ajil Jalal, Mahdi Soltanolkotabi, Eric Price, Sriram Vishwanath,
  and Alexandros~G Dimakis.
\newblock Compressed sensing with deep image prior and learned regularization.
\newblock {\em arXiv preprint arXiv:1806.06438}, 2018.

\bibitem{wang2004image}
Zhou Wang, Alan~C Bovik, Hamid~R Sheikh, Eero~P Simoncelli, et~al.
\newblock Image quality assessment: from error visibility to structural
  similarity.
\newblock {\em IEEE transactions on image processing}, 13(4):600--612, 2004.

\bibitem{whiting06pop}
Bruce~R Whiting, Parinaz Massoumzadeh, Orville~A Earl, Joseph~A O~Sullivan,
  Donald~L Snyder, and Jeffrey~F Williamson.
\newblock Properties of preprocessed sinogram data in {X}-ray computed
  tomography.
\newblock {\em Medical physics}, 33(9):3290--3303, 2006.

\bibitem{wolterink2017generative}
Jelmer~M Wolterink, Tim Leiner, Max~A Viergever, and Ivana I{\v{s}}gum.
\newblock Generative adversarial networks for noise reduction in low-dose {CT}.
\newblock {\em IEEE transactions on medical imaging}, 36(12):2536--2545, 2017.

\bibitem{yang2018low}
Qingsong Yang, Pingkun Yan, Yanbo Zhang, Hengyong Yu, Yongyi Shi, Xuanqin Mou,
  Mannudeep~K Kalra, Yi~Zhang, Ling Sun, and Ge~Wang.
\newblock Low-dose {CT} image denoising using a generative adversarial network
  with wasserstein distance and perceptual loss.
\newblock {\em IEEE transactions on medical imaging}, 37(6):1348--1357, 2018.

\bibitem{Ye2018deepR}
Dong Hye~Ye Ye, Somesh Srivastava, Jean-Baptiste Thibault, Jiang Hsieh, Ken
  Sauer, and Charles Bouman.
\newblock {Deep residual learning for model-based iterative ct reconstruction
  using plug-and-play framework}.
\newblock {\em IEEE International Conference on Acoustics, Speech and Signal
  Processing (ICASSP)}, pages 6668--6672, 2018.

\bibitem{Ye2018deep}
Jong~Chul Ye, Yoseob Han, and Eunju Cha.
\newblock Deep convolutional framelets: A general deep learning framework for
  inverse problems.
\newblock {\em SIAM Journal on Imaging Sciences}, 11(2):991--1048, 2018.

\bibitem{yokota2019dynamic}
Tatsuya Yokota, Kazuya Kawai, Muneyuki Sakata, Yuichi Kimura, and Hidekata
  Hontani.
\newblock {Dynamic PET Image Reconstruction Using Nonnegative Matrix
  Factorization Incorporated with Deep Image Prior}.
\newblock In {\em Proceedings of the IEEE International Conference on Computer
  Vision}, pages 3126--3135, 2019.

\bibitem{yoo2021time}
Jaejun Yoo, Kyong~Hwan Jin, Harshit Gupta, Jerome Yerly, Matthias Stuber, and
  Michael Unser.
\newblock Time-dependent deep image prior for dynamic mri.
\newblock {\em IEEE Transactions on Medical Imaging}, 2021.

\bibitem{yuan2020half2half}
Nimu Yuan, Jian Zhou, and Jinyi Qi.
\newblock {Half2Half: deep neural network based CT image denoising without
  independent reference data}.
\newblock {\em Physics in Medicine \& Biology}, 2020.

\bibitem{zhang2010discriminative}
Qiang Zhang and Baoxin Li.
\newblock Discriminative k-svd for dictionary learning in face recognition.
\newblock In {\em 2010 IEEE computer society conference on computer vision and
  pattern recognition}, pages 2691--2698. IEEE, 2010.

\bibitem{zhang2005total}
Xiaoqun Zhang and Jacques Froment.
\newblock Total variation based fourier reconstruction and regularization for
  computer tomography.
\newblock In {\em Nuclear Science Symposium Conference Record, 2005 IEEE},
  volume~4, pages 2332--2336. IEEE, 2005.

\bibitem{zhou2020diffraction}
Kevin~C Zhou and Roarke Horstmeyer.
\newblock Diffraction tomography with a deep image prior.
\newblock {\em Optics Express}, 28(9):12872--12896, 2020.

\end{thebibliography}
\end{document}